\documentclass[12pt,a4paper]{article}

\usepackage[]{amsmath}
\usepackage[]{amssymb}
\usepackage[]{amsfonts}
\usepackage[]{graphicx}
\usepackage[]{latexsym}
\usepackage{geometry}
\input{tcilatex}

\providecommand{\U}[1]{\protect\rule{.1in}{.1in}}

\newcommand{\be}{\begin{equation}}
\newcommand{\ee}{\end{equation}}
\newcommand{\bea}{\begin{eqnarray}}
\newcommand{\eea}{\end{eqnarray}}
\newcommand{\bean}{\begin{eqnarray*}}
\newcommand{\eean}{\end{eqnarray*}}

\def\beq{\begin{equation}}

\def\eeq{\end{equation}}

\relax

\def\spa#1{\phantom{\fbox{\rule[-#1cm]{0cm}{0cm}}}} 
\def\be{\begin{equation}}
\def\bea{\begin{eqnarray}}
\def\ee{\end{equation}}
\def\eea{\end{eqnarray}}
\def\bes{\begin{equation*}}
\def\beas{\begin{eqnarray*}}
\def\ees{\end{equation*}}
\def\eeas{\end{eqnarray*}}

\begin{document}

\vspace*{-.6in} \thispagestyle{empty}
\begin{flushright}
ROM2F/2007/11\\
LPTENS--07/27\\
\end{flushright}
\vspace{.2in}

\begin{center}
{\Large
{\bf Eikonal Approximation in AdS/CFT:}\\
\vspace{0.3cm}
{\bf Resumming the Gravitational Loop Expansion}
}

\vspace{1.5cm}

Lorenzo Cornalba$^a$, Miguel S. Costa$^{b,c}$, Jo\~ao Penedones$^b$
\vspace{1cm}

$^{a}$Dipartimento di Fisica \& INFN, Universit\'{a} di Roma ``Tor Vergata'',\\
Via della Ricerca Scientifica 1, 00133, Roma, Italy\\

\vspace{.5cm}
$^{b}$Departamento de F\'{i}sica e Centro de F\'{i}sica do Porto,\\
Faculdade de Ci\^{e}ncias da Universidade do Porto,\\
Rua do Campo Alegre, 687, 4169--007 Porto, Portugal\\

\vspace{.5cm}
$^{c}$Laboratoire de Physique Th\'eorique de l'Ecole Normale Sup\'erieure,\\
24 Rue Lhomond, 75231 Paris, France\\

\vspace{.5cm}
{\small cornalba@roma2.infn.it,
miguelc@fc.up.pt,
jpenedones@fc.up.pt}

\end{center}

\vspace{1cm}
\begin{abstract}
We derive an eikonal approximation to high energy interactions in
Anti--de Sitter spacetime, by generalizing a position space derivation of the
eikonal amplitude in flat space. We are able to resum, in terms of a
generalized phase shift, ladder and cross ladder graphs associated to the
exchange of a spin $j$ field, to all orders in the coupling constant.
Using the AdS/CFT correspondence, the resulting amplitude determines the
behavior of the dual conformal field theory four--point function $%
\left\langle \mathcal{O}_{1}\mathcal{O}_{2}\mathcal{O}_{1}\mathcal{O}%
_{2}\right\rangle $ for small values of the cross ratios, in a
Lorentzian regime. Finally we show that the phase shift is
dominated by graviton exchange and computes, in the dual CFT, the
anomalous dimension of the double trace primary operators
$\mathcal{O}_{1}\partial \cdots \partial \mathcal{O}_{2}$ of large dimension and
spin, corresponding to the relative motion of the two interacting
particles. The results are valid at strong t'Hooft coupling and 
are exact in the $1/N$ expansion.
\end{abstract}



\newpage


\section{Introduction}


In this work we pursue the program, initiated in \cite{Paper1,Paper2}, of
applying eikonal methods in the context of the AdS/CFT correspondence \cite%
{Malda1,WittenGubser,w98,Malda2,FreedmanRev}. Our main goal is to go beyond
the tree level interactions analyzed in \cite{Paper1,Paper2} and to derive
an eikonal formula for hard scattering in AdS. We shall work in the limit of
zero string length and consider the expansion in the gravitational coupling $G$.
This perturbative expansion of pure quantum gravity in AdS is dual to 
the $1/N$ expansion of gauge theories with large 't Hooft coupling $\lambda$,
since $N^2 G\sim 1$ in units of the AdS radius. 
In general, this regime of the AdS/CFT correspondence is not tractable.
On the gauge theory side, we are working at strong 't Hooft coupling. On the AdS
side, one finds the usual UV divergences of the gravitational perturbative
expansion. In this paper, we shall show that, in the particular kinematical
regime of $2\rightarrow 2$ small angle scattering at high energies, the
gravitational interaction in AdS can be resummed to all orders in $G$ using
the eikonal approximation. This amplitude determines the dual gauge theory four point
function in a particular kinematical regime and it is related
to the anomalous dimensions of  double trace primary operators with large dimension and 
spin. Although these results include all terms of the $1/N$ expansion,
there are still finite $N$ effects that are not captured by our computations.
This is the case of  instanton effects, which give rise to the usual 
non--perturbative factor $e^{-{\cal O}(1/g_s)} \sim e^{-{\cal O}(N/\lambda)}$.
Therefore we must have $N\gg \lambda $, corresponding to small string coupling $g_s\ll 1$.

We start in section \ref{eikonalflat} by rederiving the standard eikonal
approximation to ladder and cross ladder diagrams in flat space \cite{LevySucher}, 
using Feynman rules in position space. This derivation makes the physical meaning of the eikonal
approximation most transparent. Each particle follows a null
geodesic corresponding to its classical trajectory, insensitive to
the presence of the other. The leading effect of the interaction,
at large energy, is then just a phase $e^{I/4}$ determined by the tree level
interaction between the null geodesics ${\bf x}(\lambda)$ and ${\bf \bar{x}}(\bar{\lambda})$
of the incoming particles,
\begin{equation}
I= (-ig)^2 \int_{-\infty}^{\infty}d\lambda d\bar{\lambda} \,
\Pi^{(j)}\left({\bf x}(\lambda),{\bf \bar{x}}(\bar{\lambda})\right)\ ,
\label{geoint}
\end{equation}
where $g$ is the coupling and $\Pi^{(j)}$ is the propagator for the exchanged spin 
$j$ particle contracted with the external momenta. We shall see in section \ref{eikonalAdS} that this
intuitive description generalizes to AdS, resuming therefore ladder
and cross ladder Witten diagrams.

In section \ref{sectCFT} we shall explore the consequences of the eikonal approximation
in AdS for the CFT four point correlator 
\[
\hat{A}\left( \mathbf{p}_{1},\cdots ,\mathbf{p}_{4}\right) =\left\langle
\mathcal{O}_{1}\left( \mathbf{p}_{1}\right) \mathcal{O}_{2}\left( \mathbf{p}%
_{2}\right) \mathcal{O}_{1}\left( \mathbf{p}_{3}\right) \mathcal{O}%
_{2}\left( \mathbf{p}_{4}\right) \right\rangle \ ,
\]
of primary operators $\mathcal{O}_{1}$ and $\mathcal{O}_{2}$.
Using the eikonal approximation in AdS, we establish the behavior
of $\hat{A} $ in the limit of $\mathbf{p}_{1}\sim \mathbf{p}_{3}$. 
The relevant limit is not controlled by the
standard OPE, since the eikonal kinematics is intrinsically
Lorentzian. Nonetheless, the amplitude $\hat{A}$ is related to the
usual Euclidean correlator $A$ by analytic continuation and can be
easily expressed in terms of the impact parameter representation 
introduced in \cite{Paper2}.

\begin{figure}
[ptb]
\begin{center}
\includegraphics[width=8cm]{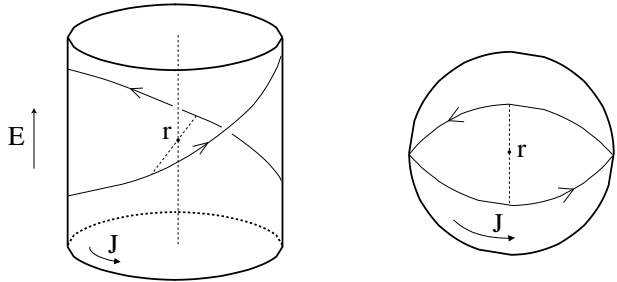}
\caption{\small Classical null trajectories of two incoming particles
moving in AdS$_{d+1}$ with total energy $E$ and
relative angular momentum $J$. They reach a minimal impact parameter $r$ is
given by $ \tanh \left( r/2\right) =J/E$.}
\label{geodesics}
\end{center}
\end{figure} 

Finally, still in section \ref{sectCFT}, we use the relation between $\hat{A}$ and $A$ to
study the conformal partial wave expansion of the Euclidean correlator $A$
in the dual channel $\mathbf{p}_{1}\rightarrow \mathbf{p}_{2}$. In this
channel, the amplitude is dominated, as in flat space, by composite states
of the two incoming particles, which are dual to specific composite primary
operators $\mathcal{O}_{1}\partial \cdots \partial \mathcal{O}_{2}$ of
classical dimension $E$ and spin $J$. We show that the eikonal approximation to $\hat{A}$
controls the anomalous dimension $2\Gamma \left( E,J\right) $ of these
intermediate two--particle states, in the limit of large $E,J$. Heuristically,
the basic idea can be summarized in two steps. Firstly, the two incoming particles
approximately follow two null geodesics in AdS$_{d+1}$ with total energy $E$ and
relative angular momentum $J$, as described by Figure \ref{geodesics}. 
The corresponding  $(d-1)$--dimensional impact parameter space is the  transverse hyperboloid $H_{d-1}$
and the minimal geodesic distance $r$ between the null geodesics is given by
$$
\tanh \left( \frac{r}{2}\right) =\frac{J}{E}\ .
$$
Then, the eikonal approximation determines the phase  $e^{-2\pi i \Gamma}$
due to the exchange of a particle of spin $j$ and dimension $\Delta$ in AdS.
As described above, this phase shift is determined by the interaction between
the two geodesics. We shall see that computing (\ref{geoint}) in AdS gives
\begin{equation}
2\,\Gamma (E,J)\simeq 
-\frac{g^{2}}{2\pi }\,(E^{2}-J^{2})^{\,j-1}\,\Pi_{\perp }\left( r\right) \ 
\ \ \ \ \ \ \ \ \ \ \ \ \ \left( E\sim J\rightarrow \infty \right)   \ ,
\label{finalres}
\end{equation}
where $g$ is the coupling in AdS and $\Pi _{\perp }$ is the Euclidean scalar
propagator of dimension $\Delta -1$ in the transverse space $H_{d-1}$.
Secondly, the phase shift is related to the anomalous dimension by the following 
argument. Recall that \cite{Malda2}, due to the
conformal structure of AdS, wave functions have
discrete allowed frequencies. More precisely, a state of
dimension $\delta$ with only positive frequencies will be almost periodic in global time $\tau $, 
acquiring only a phase $e^{-2\pi i \delta }$ as $\tau \rightarrow \tau +2\pi $. 
Since the interaction between the two particles occurs in a global time span of
$\pi$ we conclude that the full dimension of the composite state is
$\delta= E + 2\Gamma \left( E,J\right) $.

As in flat space, we deduce that the
leading contribution to $\Gamma $, for $E\sim J\rightarrow \infty $, is determined
completely by the tree level interaction, so that (\ref{finalres}) is exact
to all orders in the coupling $g$. Moreover, in gravitational theories, the
leading contribution to $\Gamma $ comes from the graviton \cite{tHooft}, with $j=2$ and 
$\Delta =d$. The result (\ref{finalres}) is then valid to all orders in the
gravitational coupling $G=g^{2}/8\pi$. For example, in the particular case 
of the duality between strings on AdS$_5\times S^5$ and four dimensional ${\cal N}=4$ SYM, 
the anomalous dimension of the above double trace operators is
$$
2\,\Gamma (E,J)\simeq - \frac{1}{4N^2}\,\frac{(E-J)^4}{EJ} \ 
\ \ \ \ \ \ \ \ \ \ \ \ \ \left( E\sim J\rightarrow \infty \right) \ ,
$$
for $E-J \ll J$ so that the impact parameter $r$ is much larger than the $S^5$ radius $\ell=1$
and the effects of massive KK modes are neglegible.

We conclude in section \ref{futurework} by briefly describing the extensions of
the results of this work to include string effects, which will
appear in a forthcoming publication \cite{Paper4}, together with
open problems and directions of future research.


\section{Eikonal Approximation in Position Space}
\label{eikonalflat}

In this section we shall rederive the standard eikonal amplitude for high energy
scattering in  Minkowski spacetime from a position space perspective.
This will prove useful because the physical picture here developed will generalize to scattering in AdS.
We shall consider $(d+1)$--dimensional Minkowski space $\mathbb{M}^{d+1}$ in close analogy with AdS$_{d+1}$.
At high energies
$$
s=(2\omega)^2
$$
we can neglect the masses of the external particles and, for
simplicity, we shall consider first an interaction mediated by a
scalar field of mass $m$. In flat space we may choose the external
particle wave functions to be plane waves $\psi_i({\bf x})=e^{\,i\,
{\bf k}_i \cdot {\bf x}}\ (i=1,\cdots,4)$, so that the amplitude
is a function of the Mandelstam invariants
$$
s=-({\bf k}_1+{\bf k}_2)^2 \ , \ \ \ \ \ \ \ \ \ \ \
t =-({\bf k}_1+{\bf k}_3)^2  =  - {\bf q}^2 \ ,
$$
We then have
$$
-2{\bf k}_1\cdot {\bf k}_2=(2\omega)^2\, ,\ \ \ \ \ \ \ \ \ \ \ \ \ {\bf k}_i^{\,2}=0\, .
$$
The eikonal approximation is valid for $s \gg -t$, where the
momentum transferred ${\bf q}={\bf k}_1+{\bf k}_3$ is
approximately orthogonal to the external momenta.

The momenta of the incoming particles naturally decompose spacetime as $\mathbb{M}^2 \times \mathbb{R}^{d-1}$.
Using coordinates $\{u,v\}$ in  $\mathbb{M}^2$ and  ${\bf w}$ in the transverse space $\mathbb{R}^{d-1} $,
a generic point can be written using the exponential map
\begin{equation}
{\bf x}=e^{\,v  {\bf T}_2 + u  {\bf T}_1 }\,  {\bf w}= {\bf w} +  u\,{\bf T}_1 + v\,{\bf T}_2\ ,
\label{coord}
\end{equation}
where the vector fields ${\bf T}_1$ and  ${\bf T}_2$ are defined by
$$
{\bf T}_1 =  \frac{{\bf k}_1 }{2\omega }\ , \ \ \ \ \ \ \ \ \
{\bf T}_2 =  \frac{{\bf k}_2}{ 2\omega}\ .
$$
The incoming wave functions are then
$$
\psi_1({\bf x})=e^{-i\omega v} \ ,\ \ \ \ \ \ \ \
\psi_2({\bf x})=e^{-i\omega u}\ .
$$
The coordinate $u$ is an affine parameter along the null geodesics
describing the classical trajectories of particle 1. This set of
null geodesics, labeled by  $v$ and  ${\bf w}$, is then the
unique congruence associated with particle 1 trajectories. Since
${\bf T}_2 = \frac{d\,}{dv}$ is a Killing vector field, these
geodesics have a conserved charge $-{\bf T}_2 \cdot {\bf k}_1=\omega$. 
At the level of the wave function this charge
translates into the condition
$$
 \mathcal{L}_{{\bf T}_2 } \psi_1 = - i\omega \psi_1 \ .
$$
Notice also that the wave function $\psi_1$ is constant along each geodesic of the
null congruence,
$$
{\bf x}(\lambda) = {\bf y} + \lambda {\bf k}_1\ ,
$$
where ${\bf k}_1=2\omega\frac{d\,}{du}$ is the momentum vector field associated to particle 1 trajectories.
Hence
$$
\mathcal{L}_{{\bf k}_1 } \psi_1 = 0\ .
$$
Finally, the field equations imply that $\psi_1$ is
independent of the transverse space coordinate ${\bf w}$.
Similar comments apply to particle 2.

Neglecting terms of order $-t/s$, the outgoing wave functions
for particles 1 and 2 are still independent of the corresponding affine parameter,
but depend on the transverse coordinate  ${\bf w}$,
$$
\psi_3({\bf x})\simeq e^{\,i\omega v+i{\bf q}\cdot {\bf w} } \ ,\ \ \ \ \ \ \ \
\psi_4({\bf x})\simeq e^{\,i\omega u-i{\bf q}\cdot {\bf w} } \ .
$$
The dependence in transverse space is determined by the
transferred momentum ${\bf q}$. Physically, the transverse space
is the impact parameter space. In fact, for two null geodesics
associated to the external particles 1 and 2, labeled
respectively by $\{v,{\bf w}\}$ and $\{\bar{u},{\bf \bar{w}}\}$,
the classical impact parameter is given by the distance  $|{\bf
w}-{\bf \bar{w}}|$.

\begin{figure} 
[ptb]
\begin{center}
\includegraphics[width=12cm]{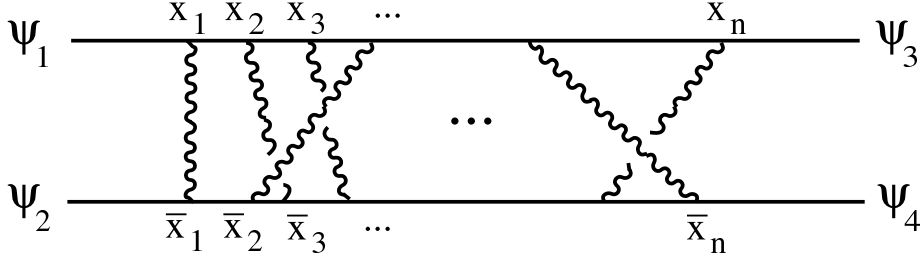}
\caption{\small The crossed--ladder graphs describing  the \emph{T}--channel exchange of many soft particles
dominate the scattering amplitude in the eikonal regime. }
\label{fig1}
\end{center}
\end{figure}

The exchange of $n$ scalar particles described by Figure \ref{fig1} gives the following contribution to
the scattering amplitude
\begin{eqnarray*}
&&\mathcal{A}_n = \frac{(-ig)^{2n}}{V}
\int_{\mathbb{M}^{d+1} } d{\bf x}_1 \cdots d{\bf x}_n
  d{\bf \bar{x}}_1 \cdots d{\bf \bar{x}}_n\,
\psi_3( {\bf x}_n )  \Delta({\bf x}_n-{\bf x}_{n-1})\cdots \Delta({\bf x}_2-{\bf x}_1)\psi_1( {\bf x}_1 )
\\
&&\ \ \ \ \ \ \psi_4( {\bf \bar{x}}_n ) \Delta({\bf \bar{x}}_n-{\bf \bar{x}}_{n-1})\cdots
\Delta({\bf \bar{x}}_2-{\bf \bar{x}}_1)\psi_2( {\bf \bar{x}}_1 )
\sum_{{\rm perm} \ {\sigma}}   \Delta_{m}({\bf x}_1-{\bf \bar{x}}_{\sigma_1})\cdots \Delta_{m}({\bf x}_n-{\bf \bar{x}}_{\sigma_n})\ ,
\end{eqnarray*}
where $V$ is the spacetime volume, $g$ is the coupling and where $\Delta({\bf x} )$ and  $\Delta_m({\bf x} )$ are,
respectively, the massless and massive Feynman propagators satisfying
$$
\left( \Box - m^2 \right) \Delta_m({\bf x} ) = i \delta({\bf x} )\ .
$$
The basic idea of the eikonal approximation is to put the horizontal propagators in Figure  \ref{fig1}
almost on--shell. This is usually done in momentum space. For example, for the propagator between vertices
${\bf x}_j$ and ${\bf x}_{j+1}$, we approximate
$$
\frac{-i}{({\bf k}_1+{\bf K})^2-i\epsilon}\simeq \frac{-i}{2{\bf k}_1 \cdot {\bf  K} -i\epsilon}\ ,
$$
where ${\bf K}$ is the total momentum transferred up to the vertex
at ${\bf x}_j $. The physical meaning of this approximation
becomes clear in the coordinates (\ref{coord}),
\begin{eqnarray}
\Delta({\bf x}_{j+1}-{\bf x}_j)&\simeq&-i
 \int \frac{ d{\bf  K}}{ (2\pi)^{d+1}}
\frac{e^{\,i({\bf k}_1+ {\bf K})\cdot ({\bf x}_{j+1}-{\bf x}_j)} }{2{\bf k}_1 \cdot {\bf K} -i\epsilon}
\nonumber \\
&\simeq&\frac{1}{2\omega}\,\Theta(u_{j+1}-u_j)\,\delta(v_{j+1}-v_j)\,\delta^{d-1}({\bf w}_{j+1}-{\bf w}_j)\ .
\label{posprop}
\end{eqnarray}
In words, particle 1 can propagate from ${\bf x}_j$ to ${\bf x}_{j+1}$ only if ${\bf x}_{j+1}$ lies on the future
directed null geodesic that starts at ${\bf x}_j$ and has tangent vector ${\bf k}_1$.
This intuitive result can be derived directly in position space. In fact, in coordinates (\ref{coord}),
the propagator satisfies
$$
\Box \Delta(x)=\left(-4\partial_u \partial_v   + \partial^2_{{\bf w} } \right) \Delta(u,v,{\bf w} )
= 2i\delta(u)\delta(v)\delta^{d-1}({\bf w})\ .
$$
Since for particle 1 we have $ \partial_v =-i\omega$, for high energies
$\Box\simeq 4i\omega\partial_u$ and (\ref{posprop}) follows.

The eikonal approximation to the position space propagators greatly simplifies the scattering
amplitude for the exchange of $n$ scalar particles
\begin{eqnarray*}
V \mathcal{A}_n \simeq
\int_{\mathbb{M}^{d+1} } d{\bf x}_1  d{\bf \bar{x}}_1
\int_{u_1}^{\infty} du_2 \int_{u_2}^{\infty} du_3 \cdots \int_{u_{n-1}}^{\infty} du_n
\int_{\bar{v}_1}^{\infty} d\bar{v}_2 \int_{\bar{v}_2}^{\infty} d\bar{v}_3 \cdots
\int_{\bar{v}_{n-1}}^{\infty} d\bar{v}_n&&\\
(4\omega)^2\left(\frac{ig}{4\omega}\right)^{2n}
 e^{\, i{\bf q}\cdot {\bf w}}\,e^{-i{\bf q}\cdot{\bf \bar{w}}}
\sum_{{\rm perm}\ {\sigma}}   \Delta_{m}({\bf x}_1-{\bf \bar{x}}_{\sigma_1})\cdots \Delta_{m}({\bf x}_n-{\bf \bar{x}}_{\sigma_n})\ ,&&
\end{eqnarray*}
with
$$
{\bf x}_j={\bf w} + u_j \,{\bf T}_1 + v\,{\bf T}_2\ ,\ \ \ \ \ \ \ \ \ \
{\bf \bar{x}}_j={\bf \bar{w}}  + {\bar u} \,{\bf T}_1 + {\bar v}_j\,{\bf T}_2\ .
$$
Furthermore, the sum over permutations can be used to extend the integrals over the affine parameters
of external particle trajectories to the full real line,
\begin{equation*}
V\mathcal{A}_n \simeq \frac{(2\omega)^2}{n!}
\int_{-\infty}^{\infty}  dv d\bar{u}
\int_{\mathbb{R}^{d-1} } d{\bf w}  d{\bf \bar{w}}\, e^{\,i{\bf q}\cdot {\bf w}}\,e^{-i{\bf q}\cdot{\bf \bar{w}}}
\left( -\frac{g^2}{16\omega^2} \int_{-\infty}^{\infty} du d\bar{v}
\,\Delta_{m}\left({\bf x}-{\bf \bar{x}} \right)\right)^n \ ,
\end{equation*}
where
$$
{\bf x}  - {\bf \bar{x}} ={\bf w} + \frac{u-{\bar u}}{2\omega}\,{\bf k}_1
-{\bf \bar{w}} - \frac{ v- \bar{v}}{2\omega}\,{\bf k}_2\ .
$$
Summing over $n$, one obtains (the $n=0$ term corresponds to the
disconnected graph)
\begin{equation}
V\mathcal{A} \simeq (2\omega)^2
\int_{-\infty}^{\infty}  dv d\bar{u}
\int_{\mathbb{R}^{d-1} } d{\bf w}  d{\bf \bar{w}}\, e^{\,i{\bf q}\cdot {\bf w}}\,e^{-i{\bf q}\cdot{\bf \bar{w}}}
\,e^{\, I/4} \ .
\label{scalareik}
\end{equation}
The integral $I$ can be interpreted as the interaction
between two null geodesics of momentum ${\bf k}_1 $ and ${\bf k}_2$
describing the classical trajectories of the incoming particles.
In fact, using as integration variables the natural affine parameters $\lambda, \bar\lambda$ along the geodesics,
one has
$$
I=(-ig)^2\int_{-\infty}^{\infty}d\lambda d\bar{\lambda}
\,\Delta_{m}\left({\bf w}+ \lambda{\bf k}_1 -{\bf \bar{w}} - \bar{\lambda} {\bf k}_2\right)\ .
$$
Explicit computation yields the Euclidean propagator $\Delta_{\perp}$ of mass $m$ in the transverse
$\mathbb{R}^{d-1}$ space
$$
I=\frac{-ig^2}{ {\bf k}_1\cdot  {\bf k}_2 }\int_{\mathbb{R}^{d-1} }\frac{d{\bf k}_{\perp}}{(2\pi)^{d-1}}\,
\frac{e^{\,i{\bf k}_{\perp}\cdot ({\bf w} -{\bf \bar{w}}) }}{{\bf k}^2_{\perp} +m^2}
=\frac{2ig^2}{s}\Delta_{\perp}( {\bf w} -{\bf \bar{w}})\ ,
$$
and we obtain the well known eikonal amplitude
\begin{equation}
\mathcal{A} (s,t=-{\bf q}^2) \simeq 2s
\int_{\mathbb{R}^{d-1} } d{\bf w}  \,
e^{\, i{\bf q} \cdot{\bf w}  +\frac{ig^2}{2s} \Delta_{\perp}\left({\bf w} \right)}  \ .
\end{equation}

The generalization of the above method to interactions mediated by a spin $j$ particle is now straightforward.
We only need to change the integral $I$ describing the scalar interaction between null geodesics.
The spin $j$ exchange alters the vertices of the local interaction, as well as the propagator
of the exchanged particles.
In general, the vertex includes $j$ momentum factors with a complicated
index structure. However, in the eikonal regime, the momentum entering the vertices
are approximately the incoming momenta ${\bf k}_1, {\bf k}_3$.
Therefore, the phase $I$ should be replaced  by
  \footnote{
The sign $(-)^j$ indicates that, for odd $j$, particles 1 and 2 have opposite charge
with respect to the spin $j$ interaction field. With this convention the interaction
is attractive, independently of $j$.}
\begin{eqnarray*}
I&=&-g^2(-2)^j\,({\bf k}_1)_{\alpha_1}\cdots ({\bf k}_1)_{\alpha_j}\,
({\bf k}_2)_{{\beta}_1}\cdots ({\bf k}_2)_{{\beta}_j}
\\
&&\int_{-\infty}^{\infty}d\lambda d\bar{\lambda}
\,\Delta_m^{\ \alpha_1 \cdots \alpha_j\,{\beta}_1 \cdots {\beta}_j }
\left({\bf w}+ \lambda{\bf k}_1 -{\bf \bar{w}} - \bar{\lambda} {\bf k}_2\right)\ ,
\end{eqnarray*}
where $\Delta_m^{\ \alpha_1 \cdots \alpha_j\,{\beta}_1 \cdots {\beta}_j }$
is the propagator of the massive spin $j$ field. Recall that the equations of motion for
a spin $j$ field $h^{\alpha_1 \cdots \alpha_j}$ imply that $h$ is symmetric, traceless
and transverse ($\partial_{\alpha_1}\,h^{\alpha_1 \cdots \alpha_j}=0$), together with the
mass--shell condition $\square=m^2$. Therefore, the relevant part of the propagator at high energies
is given by
$$
\eta^{(\alpha_1 \beta_1}\eta^{\alpha_2 \beta_2}\cdots \eta^{\alpha_j \beta_j)}\, \Delta_m\left({\bf x} -{\bf \bar{x}} \right)\ +\cdots,
$$
where the indices $\alpha_i$ and $\beta_i$ are separately symmetrized with weight $1$. The neglected terms in $\cdots$ are trace terms, 
which vanish since ${\bf k}_i^{\,2}=0$, and derivative terms acting on $\Delta_m$, which vanish after integration along the two interacting 
geodesics. Compared to the scalar case, we have then an extra factor of $(-2{\bf k}_1\cdot{\bf k}_2)^{j}=s^j$, so that
$$
I = 2 i g^2\, s^{\, j-1} \,\Delta_{\perp}( {\bf w} -{\bf \bar{w}})\ .
$$
Note that we have normalized the coupling $g^2$ so that the leading behavior of the tree level amplitude at large $s$
is given by $-g^2s^j/t$. In the particular case of $j=2$ we then have $g^2=8\pi G$,
where $G$ is the canonically normalized Newton constant.

It is known \cite{Jackiw, Kabat} that the eikonal approximation is problematic for $j=0$ exchanges. In this case, the large
incoming momentum can be exchanged by the mediating particle, interchanging the role of $u,v$ in intermediate parts of the graph.
The eikonal approximation estimates correctly the large $s$ behavior of the amplitude at each order in perturbation theory, but
underestimates the relative coefficients, which do not resum to an exponential. Nonetheless, this is not problematic, since exactly in the
$j=0$ case the higher order terms are suppressed by powers of $s^{-1}$. For $j\geq 1$ the problematic
hard exchanges are suppressed at large energies and the eikonal approximation is valid. On the other hand,
for the QED case where $j=1$, there is a different set of graphs involving virtual fermions \cite{Kabat}  that 
dominate the eikonal soft photons exchange. Therefore, also for $j=1$, the validity of the eikonal approximation 
is in question. None of these problems arise, though, for the most relevant case, the gravitational interaction with $j=2$.


\section{Eikonal Approximation in Anti--de Sitter}
\label{eikonalAdS}

Let us now apply the intuitive picture developed in the previous section
to the eikonal approximation in position space to hard scattering in
Anti--de Sitter spacetime.
Recall that AdS$_{d+1}$ space, of dimension $d+1$ and radius $\ell = 1$,
can be defined as a pseudo--sphere
in the embedding space $\mathbb{R}^{2,d}$ given by the set
of points  \footnote{Rigorously, AdS space is the universal covering of this pseudo--sphere.
We shall use units such that $\ell=1$.}
\begin{equation}
{\bf x} \in \mathbb{R}^{2,d}\ ,\ \ \ \ \ \ \ \ \
\mathbf{x}^{2}=-1\ .\label{pseudoS}
\end{equation}
In the remainder of this paper, points, vectors and scalar
products are taken in the embedding space $\mathbb{R}^{2,d}$,
except when extra care is needed with the AdS global structure 
or when we wish to make contact with the dual CFT notation.

Consider  the Feynman graph in Figure \ref{fig1}, but now  in AdS.
For simplicity, we consider the exchange of an AdS scalar field of dimension $\Delta$ and, for external fields,
we consider scalars of dimension $\Delta_1$ and $\Delta_2$. Then, the graph in Figure \ref{fig1} evaluates to
\begin{equation}
\begin{array}{c}
\displaystyle{A_n = (ig)^{2n} \int_{{\rm AdS}}d{\bf x}_1 \cdots d{\bf x}_n d{\bf \bar{x}}_1 \cdots d{\bf \bar{x}}_n \,
\psi_3( {\bf x}_n )  \Pi_{\Delta_1}({\bf x}_n,{\bf x}_{n-1})\cdots \Pi_{\Delta_1}({\bf x}_2,{\bf x}_1)\psi_1( {\bf x}_1)}
\spa{0.5}
\\
\displaystyle{\psi_4( {\bf \bar{x}}_n ) \Pi_{\Delta_2}({\bf \bar{x}}_n,{\bf \bar{x}}_{n-1})\cdots
\Pi_{\Delta_2}({\bf \bar{x}}_2,{\bf \bar{x}}_1)\psi_2( {\bf \bar{x}}_1 )
\sum_{{\rm perm} \ \sigma}   \Pi_{\Delta}({\bf x}_1,{\bf \bar{x}}_{\sigma_1})
\cdots \Pi_{\Delta}({\bf x}_n,{\bf \bar{x}}_{\sigma_n})}\ ,\label{AdSamplitude}
\end{array}
\end{equation}
where $\Pi_{\Delta}({\bf x},{\bf \bar{x}})$ stands for the scalar propagator of mass $\Delta(\Delta-d)$ in AdS, satisfying
\begin{equation}
\left[ \Box_{\rm AdS}-\Delta( \Delta-d ) \right]\Pi_{\Delta}({\bf x},{\bf \bar{x}})=i\delta({\bf x},{\bf \bar{x}})\ .
\label{propeq}
\end{equation}
In general, this amplitude is very hard to compute. However, we expect some drastic simplifications
for specific external wave functions describing highly energetic particles scattering at
fixed impact parameters. In analogy with flat space, we expect the eikonal approximation to correspond
to the collapse of the propagators $\Pi_{\Delta_1}$ and $\Pi_{\Delta_2}$ into
null geodesics approximating classical trajectories of highly energetic particles.

\subsection{Null Congruences in AdS and Wave Functions}\label{NCWF}

\begin{figure}
\begin{center}
\includegraphics[width=9cm]{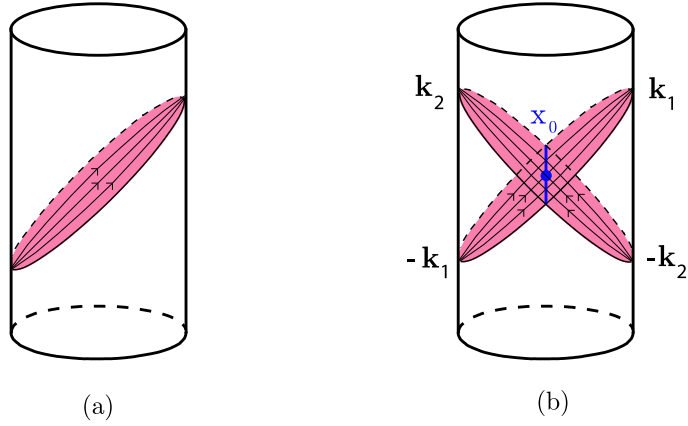}
\caption{\small (a) A generic null hypersurface  ${\bf k}
\cdot{\bf y} =0$ in conformally compactified AdS. (b) The two null
hypersurfaces  ${\bf k}_1 \cdot{\bf y} =0$ and  ${\bf k}_2
\cdot{\bf y} =0$. Their intersection is the transverse hyperboloid
$H_{d-1}$ containing the reference point ${\bf x}_0$.
We shall see in section \ref{sectCFT} that the null vectors ${\bf k}_i$ and $-{\bf k}_i$
can be thought of as points in the AdS conformal boundary. }
\label{nullsurface}
\end{center}
\end{figure}

A null geodesic in AdS is also a null geodesic in the embedding space
$$
{\bf x}(\lambda)={\bf y}+\lambda\, {\bf k} \ ,
$$
where ${\bf y}\in {\rm AdS}$ and the tangent vector ${\bf k}$ satisfies
$$
{\bf k}^2=0 \ ,\ \ \ \ \ \ \ \ \ \ \ \ \ \ {\bf k} \cdot{\bf y} =0 \ .
$$
We will follow the intuitive idea that the wave functions $\psi_1$ and $\psi_2$ correspond to the initial
states of highly energetic particles moving along two intersecting congruences of null geodesics.
As described in the previous section, in flat space there is a one--to--one correspondence between
null momenta (up to scaling) and congruences of null geodesics.
On the other hand, in AdS the situation is more complicated.
Given a null vector $ {\bf k}$ there is a natural set of null geodesics $ {\bf y}+\lambda {\bf k}$
passing through all points ${\bf y}\in {\rm AdS}$ belonging to the hypersurface ${\bf k} \cdot{\bf y} =0$,
as shown in Figure \ref{nullsurface}(a). However,
to construct a congruence of null geodesics we need to extend this set to the full
AdS space. Contrary to flat space, in AdS this extension is not unique because
the spacetime conformal boundary is timelike. We will now describe how to construct
such a congruence in analogy with the construction presented for flat space.

We start with two null vectors ${\bf k}_1,\, {\bf k}_2$ associated with the incoming particles, 
as represented in  Figure \ref{nullsurface}(b) and normalized as in flat space  
$$-2{\bf k}_1\cdot {\bf k}_2 = (2\omega)^2\  .$$
The transverse space is naturally defined as the intersection of the two null hypersurfaces associated to 
${\bf k}_1$ and ${\bf k}_2$.  It is the  hyperboloid $H_{d-1}$ defined by
$$
{\bf w} \in {\rm AdS} \ , \ \ \ \ \
{\bf k}_1 \cdot{\bf w} ={\bf k}_2 \cdot{\bf w} =0\ .
$$
In order to introduce coordinates in AdS$_{d+1}$ in analogy with (\ref{coord}), we choose an arbitrary
reference point ${\bf x}_0$ in this transverse space $H_{d-1}$. This allows us to define the vector fields
$$
{\bf T}_1 ({\bf x})  = 
\frac{({\bf k}_1\cdot {\bf x})\, {\bf x}_0 -  ({\bf x}_0 \cdot {\bf x})\, {\bf k}_1}{2\omega}
   \ , \ \ \ \ \ \ \ \ \
{\bf T}_2({\bf x})   = 
\frac{({\bf k}_2\cdot {\bf x})\, {\bf x}_0 -  ({\bf x}_0 \cdot {\bf x})\, {\bf k}_2}{2\omega}  \ ,
$$
which, from the embedding space perspective, are respectively the
generators of parabolic Lorentz transformations in the ${\bf
x}_0\,{\bf k}_1$ and   ${\bf x}_0\,{\bf k}_2$--plane. They
therefore generate AdS isometries. We may now introduce
coordinates $\{u,v,{\bf w}\}$ for ${\bf x} \in $ AdS$_{d+1}$ as
follows
\begin{equation}
\begin{array}{rcl}
{\bf x}&=& \displaystyle{e^{\,v  {\bf T}_2}\,e^{\,u  {\bf T}_1}\, {\bf w}}\spa{0.2}\\
&=& \displaystyle{{\bf w} -  u\, \frac{({\bf x}_0\cdot{\bf w}) \, {\bf k}_1}{2\omega} - v\,\frac{({\bf x}_0\cdot{\bf w}) \,{\bf k}_2}{2\omega}
+ uv\,\frac{({\bf x}_0\cdot{\bf w}) \,{\bf x}_0}{2} +  uv^2\,\frac{({\bf x}_0\cdot{\bf w}) \,{\bf k}_2}{8\omega}} \ ,
\end{array}
\label{coordAdS}
\end{equation}
where ${\bf w}\in H_{d-1}$.
It is important to realize that, contrary to the flat space case,  
$[{\bf T}_1, {\bf T}_2]\neq 0$ and therefore the order of the exponential maps in (\ref{coordAdS}) is important, as will become clear below.

As for flat space, the coordinate $u$ is an affine parameter along
null geodesics labeled by $v$ and ${\bf w}$, which form the
desired congruence for particle 1. In fact,  (\ref{coordAdS}) can
be written as
$$
{\bf x}= e^{\,v  {\bf T}_2}\, {\bf w} +u\, e^{\,v  {\bf T}_2} \, {\bf T}_1 ({\bf w} )\ .
$$
Hence, the geodesics in the null congruence associated to particle 1 are given by
\begin{equation}
{\bf x}={\bf y}+\lambda\, {\bf k}\, ,
\label{p}
\end{equation}
where
$$
{\bf y} = e^{\,v  {\bf T}_2}\, {\bf w} =
{\bf w}  - v\,\frac{({\bf x}_0\cdot{\bf w}) \,{\bf k}_2}{2\omega}\, .
$$
The normalization of the momentum  ${\bf k}$ and affine parameter $\lambda$ of the classical trajectories is fixed by
demanding, as in flat space, that the conserved charge 
$-{\bf T}_2\cdot {\bf k}=\omega$.
This gives
\begin{eqnarray*}
\lambda &=&  u\,\frac{({\bf x}_0\cdot{\bf w})^2}{2\omega}\, ,\\
{\bf k} &=& \frac{2\omega}{({\bf x}_0\cdot{\bf w})^2  } \,e^{\,v  {\bf T}_2}\, {\bf T}_1({\bf w} )
= \frac{2\omega}{({\bf x}_0\cdot{\bf w})^2}\,\frac{d\,}{du}
=- \frac{1}{{\bf x}_0\cdot{\bf w}  } \left(
{\bf k}_1-v\omega \,{\bf x}_0-\frac{v^2}{4}\,{\bf k}_2
\right)\ .
\end{eqnarray*}
Let us we remark that different choices of ${\bf x}_0$ give different congruences, all containing
the null geodesics ${\bf w}+\lambda '\, {\bf k}_1$, which lye on
the  hypersurface ${\bf k}_1\cdot {\bf x} = 0$ at $v=0$. Starting from this hypersurface,
we  then constructed a congruence of null geodesics using
the AdS isometry generated by ${\bf T}_2$.

\begin{figure}
\begin{center}
\includegraphics[width=14cm]{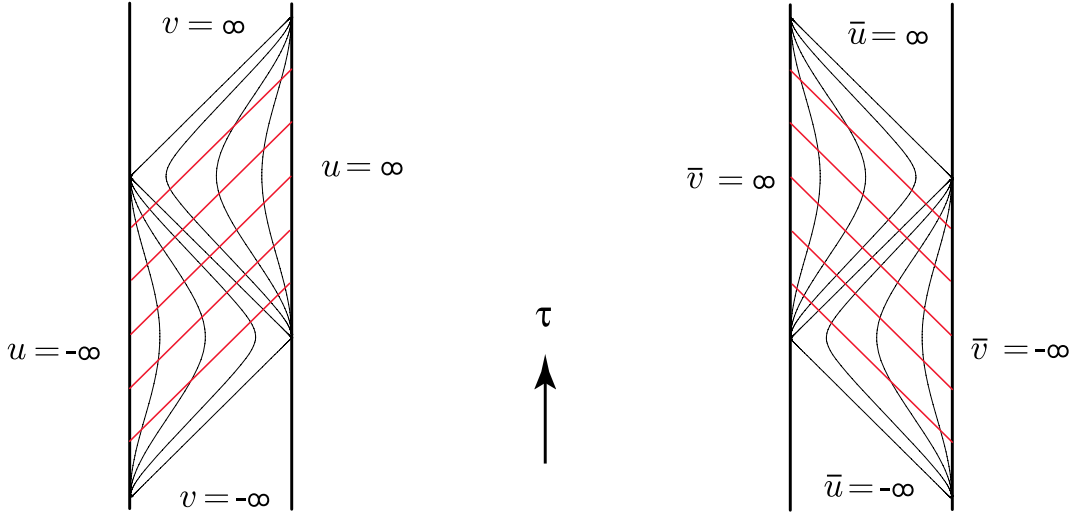}
\caption{\small The coordinates $\{u,v\}$ and $\{\bar{u},\bar{v}\}$ for the simplest case of AdS$_2$.
In general, the wave function of particle 1 is independent of the coordinate $u$, while that of particle
2 is independent of the coordinate $\bar{v}$.}
\label{uvcoord}
\end{center}
\end{figure}

Contrary to flat space, the curves defined by constant $u$ and ${\bf w}$ in the coordinate system
(\ref{coordAdS}) are {\em not} null geodesics
(except for the curves on the surface $u=0$ which are null geodesics with affine parameter $v$).
These curves are the integral curves of the Killing vector field $ {\bf T}_2 = \frac{d\,}{dv}$.
In fact, these curves are not even null, as can be seen from the form of the AdS metric in these coordinates
\begin{equation}
ds^2=d {\bf w}^2- ({\bf x}_0\cdot{\bf w}  )^2 dudv - \frac{u^2}{4} ({\bf x}_0\cdot{\bf w}  )^2   dv^2\ ,
\label{AdSMetric}
\end{equation}
where $d {\bf w}^2$ is the metric on the hyperboloid $H_{d-1}$.
To construct the null congruence for particle 2 we introduce new coordinates
$\{\bar{u} ,   \bar{v},   {\bf \bar{w}} \}$ for ${\bf \bar{x}} \in $ AdS$_{d+1}$ as follows
$$
{\bf \bar{x}}= e^{\,\bar{u}  {\bf T}_1}\,e^{\,\bar{v}  {\bf T}_2}\, {\bf \bar{w}}\ .
$$
The two sets of coordinates are related by
\beas
&&\bar{u}=u\left(1-\frac{uv}{4}\right)^{-1}\\
&&\bar{v}=v\left(1-\frac{uv}{4}\right)\\
&&\bar{\bf w}={\bf w}
\eeas
The congruence associated with particle 2 is then the set of null geodesics
\begin{equation}
{\bf \bar{x}}={\bf \bar{y}}+\bar{\lambda}\, {\bf \bar{k}} \ ,
\label{pbar}
\end{equation}
with
\begin{eqnarray*}
&&{\bf \bar{y}}= e^{\,\bar{u}  {\bf T}_1}\, {\bf \bar{w}}=
{\bf \bar{w}}  - \bar{u}\,\frac{({\bf x}_0\cdot{\bf \bar{w}}) \,{\bf k}_1}{2\omega}\ ,\\
&&\bar{\lambda} =  \bar{v}\,\frac{({\bf x}_0\cdot{\bf \bar{w}})^2}{2\omega} \ ,\\
&&{\bf \bar{k}}= \frac{2\omega}{({\bf  x}_0\cdot{\bf  \bar{w}})^2 }\,
e^{\, \bar{u}  {\bf T}_1}\, {\bf T}_2({\bf  \bar{w}} )
=\frac{2\omega}{({\bf x}_0\cdot{\bf \bar{w}})^2}\,\frac{d\,}{d\bar{v}}
= - \frac{1}{{\bf x}_0\cdot {\bf  \bar{w}}  } \left(
{\bf k}_2- \bar{u}\omega \,{\bf x}_0-\frac{\bar{u}^2}{4}\,{\bf k}_1
\right)\ ,
\end{eqnarray*}
so that the conserved charge $-{\bf T}_1\cdot {\bf \bar{k}}=\omega$.
In Figure \ref{uvcoord} we plot the curves of constant $u$ and $v$ (left) and of constant $\bar{u}$ and $\bar{v}$ (right)
in the simplest case of AdS$_2$.

As in flat space, the wave function describing particle 1 carries energy $\omega$
$$
\mathcal{L}_{{\bf T}_2} \psi_1 = \partial_v \psi_1 \simeq - i\omega \psi_1 \ .
$$
Therefore we choose
$$
\psi_1( {\bf x})= e^{-i\omega  v}F_1({\bf x} )\ ,
$$
where the function $F_1$ is approximately constant over the length scale $1/\omega$,
more precisely $|\partial F_1| \ll \omega |F_1|$.
The Klein--Gordon equation for the wave function $\psi_1$ implies
$$
\left[ \frac{4i\omega}{({\bf x}_0\cdot{\bf w})^2}\, \partial_u +\Box_{\rm AdS}- \Delta_1(\Delta_1-d) \right]F_1({\bf x} )=0\ ,
$$
since, as in flat space, the coordinate $v$ satisfies
$$
\square_{\rm AdS}\ v = (\nabla v)^2 = 0\, .
$$
The above equation can then be solved expanding $F_1$ in powers of $1/\omega$,
$$
F_1({\bf x} )=F_1(v,{\bf w})- \frac{({\bf x}_0\cdot{\bf w})^2}{4i\omega}
\int du \Big[\Box_{\rm AdS}-\Delta_1(\Delta_1-d) \Big]F_1(v,{\bf w})+\cdots
$$
Since the eikonal approximation gives only the leading behavior of the scattering amplitude at large
$\omega$, it is enough to consider only the first term
$F_1({\bf x} )\simeq F_1(v,{\bf w})$ so that,
to this order, we have
$$
\mathcal{L}_{\bf k} \psi_1 = 0\ ,
$$
as expected. We conclude that the function $F_1$ is a smooth transverse modulation independent of the affine parameter $\lambda$ of
the null geodesics associated with the classical trajectories of particle 1.
Similar reasoning applied to particle 2 leads to
$$
\psi_2( {\bf \bar{x}})\simeq e^{-i\omega \bar{u}}F_2(\bar{u},{\bf \bar{w}})\ .
$$

Finally, since in the eikonal regime the particles are only slightly deviated by the scattering process,
to leading order in $1/\omega$ the outgoing wave functions
are also independent of the corresponding affine parameters,
$$
\psi_3( {\bf x})\simeq e^{i\omega v}F_3(v, {\bf w})\ ,\ \ \ \ \ \ \ \
\psi_4( {\bf \bar{x}})\simeq e^{i\omega \bar{u}}F_4(\bar{u} , {\bf \bar{w}})\ ,
$$
with the same requirement $|\partial F| \ll \omega |F|$.

We have kept the discussion of this section completely coordinate
independent. On the other hand, given the choice of ${\bf k}_1$ and ${\bf k}_2$, the embedding
space ${\mathbb R}^{2,d}$ naturally splits into $\mathbb{M}^{2}\times \mathbb{M}^{d}$,
with $\mathbb{M}^{2}$ spanned by ${\bf k}_1$ and $ {\bf k}_2$ and with $\mathbb{M}^{d}$ its orthogonal complement.
We may then introduce coordinates ${\bf x} = (x^+,x^-,x^a)$ where $x^\pm$ are 
light-cone coordinates on $\mathbb{M}^{2}$ and where the $x^a$ parametrize $\mathbb{M}^{d}$. 
We shall often omit the explicit label $a$. AdS is then given by
$$
{\bf x}^2 = -x^+ x^- + x\cdot x = -1\,.
$$
In terms of these coordinates we have that
$$
{\bf k_1} = -2\omega\, (0,1,0)\ ,  \ \ \ \ \ \ \ \ \ {\bf k_2} = -2\omega\, (1,0,0) 
$$
and
\begin{equation}
{\bf x_0} = (0,0,x_0) \ , \ \ \ \ \ \ \ \ {\bf w} = (0,0,{\rm w})\ ,\ \ \ \ \ \ \ \ 
 {\bf \bar{w}} = (0,0,{\rm \bar{w}})\ .
\label{x0}
\end{equation}
Then, the vector fields ${\bf T}_1$ and $ {\bf T}_2$ are simply the parabolic Poincar\'e translations
\beas
&&{\bf T}_1=\frac{1}{2}\,x^+ (x_0\cdot\partial)+(x_0\cdot x)\,\partial_-\, ,\\
&&{\bf T}_2=\frac{1}{2}\,x^- (x_0\cdot\partial)+(x_0\cdot x)\,\partial_+\, .
\eeas
The action of $e^{\, \alpha {\bf T}_1}$ is simple in the Poincar\'e parametrization ${\bf x} =1/r\,(1,r^2+y^2,y)$ and 
corresponds to translations $y \to y+(\alpha /2)\,  x_0$ in the time direction indicated by $x_0$. This corresponds to
\beas
&&x^- \to x^- + \alpha\, x_0\cdot x + \frac{\alpha^2}{4}\, x^+\ ,\\
&&x   \to x + \frac{\alpha}{2}\, x^+ x_0\ ,
\eeas
with $x^+$ fixed. Similar remarks apply to ${\bf T}_2$ with the roles of $x^+$ and $x^-$ interchanged.

\subsection{Eikonal Amplitude}

We are now in position to compute the leading behavior of the amplitude (\ref{AdSamplitude})
for the exchange of $n$ scalars in AdS at large $\omega$, using the techniques explained in section 2.
The first step is to obtain an approximation for the AdS propagator similar to (\ref{posprop}).
Since for particle 1 we have $\partial_v \simeq -i\omega $, we can approximate
$$
\Box_{\rm AdS} \simeq \frac{4i\omega}{({\bf x}_0\cdot{\bf w})^2} \,\partial_u\ ,
$$
in equation (\ref{propeq}) for the propagator of particle 1 between vertices ${\bf x}_j$ and  ${\bf x}_{j+1}$,
obtaining
$$
\frac{4i\omega}{({\bf x}_0\cdot{\bf w})^2} \,\partial_{u_j} \Pi_{\Delta_1}({\bf x}_j,{\bf x}_{j+1})=
\frac{2i}{({\bf x}_0\cdot{\bf w})^2}\,\delta(u_j-u_{j+1})\, \delta(v_j-v_{j+1})\,\delta_{H_{d-1}}({\bf w}_j,{\bf w}_{j+1})\ .
$$
The solution,
$$
\Pi_{\Delta_1}({\bf x}_j,{\bf x}_{j+1})\simeq \frac{1}{2\omega}\,\Theta(u_j-u_{j+1})\,
\delta(v_j-v_{j+1})\,\delta_{H_{d-1}}({\bf w}_j,{\bf w}_{j+1})\ ,
$$
has the natural interpretation of propagation only along the particle classical trajectory and, in these
coordinates, takes almost exactly the same form as the corresponding propagator (\ref{posprop}) in flat space.
With this approximation to the propagator, the amplitude (\ref{AdSamplitude}) associated with
the exchange of $n$ scalar particles simplifies to
\begin{eqnarray*}
A_n &\simeq& (2\omega)^2 \int_{-\infty}^{\infty}dv d\bar{u} \int_{H_{d-1}} d{\bf w} d{\bf \bar{w}}
 F_1(v,{\bf w}) F_3(v,{\bf w})F_2(\bar{u} , {\bf \bar{w}})F_4(\bar{u} , {\bf \bar{w}})\\
&&
\int_{-\infty}^{\infty} du_1 \int_{u_1}^{\infty} du_2  \cdots \int_{u_{n-1}}^{\infty} du_n
\int_{-\infty}^{\infty} d\bar{v}_1\int_{\bar{v}_1}^{\infty} d\bar{v}_2  \cdots
\int_{\bar{v}_{n-1}}^{\infty} d\bar{v}_n\\
&&\left(\frac{ig \,({\bf x}_0\cdot{\bf w})( {\bf x}_0\cdot  {\bf \bar{w}}) }{4\omega}\right)^{2n}
\sum_{{\rm perm}\ \sigma}   \Pi_{\Delta}({\bf x}_1,{\bf \bar{x}}_{\sigma_1})\cdots
 \Pi_{\Delta}({\bf x}_n,{\bf \bar{x}}_{\sigma_n})\ ,
\end{eqnarray*}
where
$$
{\bf x}_j = e^{\,v   {\bf T}_2 }\,e^{\,u_j  {\bf T}_1}\, {\bf w} \ ,\ \ \ \ \ \ \ \
{\bf \bar{x}}_j= e^{\,\bar{u}  {\bf T}_1 }\,e^{\,\bar{v}_j  {\bf T}_2 }\, {\bf \bar{w}}\ .
$$
Notice that the extra powers of $({\bf x}_0\cdot{\bf w})^2/2$ and $( {\bf x}_0\cdot  {\bf \bar{w}})^2/2$ come from the integration
measure in the ${\bf x}_j$ and ${\bf \bar{x}}_j$ coordinates, respectively.
As for flat space, the integrals over the affine parameters can be extended
to the real line so that, after summing over $n$, we obtain
\begin{equation}
A \simeq (2\omega)^2 \int_{-\infty}^{\infty}dv d\bar{u} \int_{H_{d-1}} d{\bf w} d{\bf \bar{w}}
 F_1(v,{\bf w}) F_3(v,{\bf w})F_2(\bar{u} , {\bf \bar{w}})F_4(\bar{u} , {\bf \bar{w}})\, e^{\,I/4}\ ,
\label{eqFinal}
\end{equation}
with
$$
I= -\frac{g^2 \,({\bf x}_0\cdot{\bf w})^2 ( {\bf x}_0\cdot  {\bf \bar{w}})^2 }{(2\omega)^2}
\int_{-\infty}^{\infty} du d\bar{v} \, \Pi_{\Delta}\left({\bf x},{\bf \bar{x}}\right)\ .
$$
This can be rewritten as the tree--level interaction between two classical trajectories of the incoming particles
described by (\ref{p}) and (\ref{pbar}),
which are labeled respectively by ${\bf y}$  and ${\bf \bar{y}}$,
$$
I= (-ig)^2\,\int_{-\infty}^{\infty}d\lambda d\bar{\lambda}\,
\Pi_{\Delta}\left({\bf y}+ \lambda  \, {\bf k}({\bf y}) ,
{\bf \bar{y}} + \bar{\lambda} \, {\bf \bar{k}}({\bf \bar{y}}) \right)\ .
$$
Hence, the AdS eikonal amplitude just obtained is the direct analogue of the corresponding
flat space amplitude (\ref{scalareik}).

The generalization of the above result to the case
of interactions mediated by a minimally coupled particle of spin $j$ is straightforward,
and we shall only give the relevant results. At high energies, the only
change concerns the propagator $\Pi_\Delta$, which now should be replaced by the propagator
of the spin--$j$ particle contracted with the null momenta of the geodesics
$$
\Pi^{(j)}_\Delta=(-2)^j\, {\bf k}_{\alpha_1}\cdots{\bf k}_{\alpha_j}\, {\bf \bar{k}}_{\beta_1}\cdots{\bf \bar{k}}_{\beta_j}
\, \Pi_\Delta^{\alpha_1,\cdots,\alpha_j,\beta_1,\cdots,\beta_j}\, ,
$$
where the indices $\alpha_i,\beta_j$ are tangent indices to AdS.
This follows immediately from the fact that, at high energies,
covariant derivatives $-i\nabla_\alpha$ in interaction
vertices can be replaced  by ${\bf k}_\alpha$ and ${\bf \bar k}_\alpha$ for
particle one and two, respectively. The spin--$j$ propagator is
totally symmetric and traceless in the indices $\alpha_1\dots
\alpha_j$ (and similarly in the indices $\beta_1\cdots \beta_j$),
it is divergenceless and satisfies
\begin{equation}
\left[ \Box -\Delta( \Delta-d )+j \right]
\Pi_\Delta^{\alpha_1,\cdots,\alpha_j,\beta_1,\cdots,\beta_j}({\bf x},{\bf \bar{x}})=
i\,g^{(\alpha_1\beta_1}g^{\alpha_2\beta_2}\cdots g^{\alpha_1\beta_1)}\,\delta({\bf x},{\bf \bar{x}})\,+\cdots ,
\label{propeqBIS}
\end{equation}
where the indices $\alpha_i$ and $\beta_i$ are separately symmetrized, and where the terms in $\cdots$  
contain derivatives of $\delta({\bf x},{\bf \bar{x}})$ and are not going to be of relevance to the discussion 
which follows, since they give subleading contributions at high energies. The eikonal expression 
(\ref{eqFinal}) is then valid in general, with the phase factor $I$ now replaced by
$$
I= -g^2\,\int_{-\infty}^{\infty}d\lambda d\bar{\lambda}\,
\Pi^{(j)}_{\Delta}\left({\bf y}+ \lambda  \, {\bf k} ,
{\bf \bar{y}} + \bar{\lambda} \, {\bf \bar{k}} \right)\ .
$$
Note that we have normalized the interaction coupling as in flat space, where the tree level interaction is given by $-g^2\, s^{j}/t$
at large $s$.

\subsection{Transverse Propagator}\label{sectTP}

Now we compute the integral $I$. Its last expression shows that it is
a Lorentz invariant local function of ${\bf y},{\bf \bar{y}},{\bf k}$ and ${\bf \bar{k}}$. Moreover,
it is invariant under
$$
{\bf y}\to {\bf y}+\alpha\, {\bf k} \ ,\ \ \ \ \ \ \ \ \ \
{\bf \bar{y}} \to{\bf \bar{y}} + \bar{\alpha} \, {\bf \bar{k}}\ ,
$$
and it scales like $ I\to (\alpha\bar{\alpha})^{j-1}I $
when ${\bf k}\to \alpha\, {\bf k} $ and ${\bf \bar{k}} \to\bar{\alpha} \, {\bf \bar{k}} $.
Therefore, the integral $I$ is fixed up to  an undetermined function $G$,
\begin{eqnarray*}
I&=&2ig^2\,(-2{\bf k}\cdot{\bf \bar{k}})^{j-1}\,G\left({\bf y}\cdot {\bf \bar{y}}  -
\frac{({\bf k}\cdot{\bf \bar{y}}) \, ({\bf \bar{k}}\cdot{\bf y})}{{\bf k}\cdot   {\bf \bar{k}} } \right)\\
&=&2ig^2\,s^{j-1}
\,G(  {\bf w} \cdot {\bf \bar{w}} )
\ ,
\end{eqnarray*}
with $s$ defined in analogy with flat space
\begin{equation}
s=-2{\bf k}\cdot{\bf \bar{k}}=(2\omega)^{2}
\frac{(1+v\bar{u}/4)^2}{
({\bf x}_0\cdot{\bf w})\, ({\bf x}_0\cdot  {\bf \bar{w}})}\ .
\label{sAdS}
\end{equation}
To determine the function $G$ we use equation (\ref{propeqBIS}), contracting both sides with
$$
(-2)^j\, {\bf k}_{\alpha_1}\cdots{\bf k}_{\alpha_j}\, {\bf \bar{k}}_{\beta_1}\cdots{\bf \bar{k}}_{\beta_j}
$$
and integrating against
$$
\int_{-\infty}^{\infty}du d\bar{v} = \frac{(2\omega)^2}{({\bf x}_0\cdot{\bf w})^2({\bf \bar{x}}_0\cdot{\bf \bar{w}})^2}\,\int_{-\infty}^{\infty}d\lambda d\bar{\lambda}\, .
$$
Here we discuss the simplest case of $j=0$, leaving the general case to appendix \ref{app1}. Consider then first the integral of the RHS of (\ref{propeq}).
Using the explicit form of the $\delta$--function in the  $\{u,v, {\bf w} \}$ coordinate system,
$$
\delta({\bf x},{\bf \bar{x}}) =
 \frac{2}{ ({\bf x}_0\cdot{\bf w})^2}\,
\delta_{H_{d-1}}({\bf w},{\bf \bar{w}})\,
\delta \left(u-\bar{u} \left(1-\frac{ \bar{u}\bar{v}}{4}\right) \right)\,
\delta \left(v- \bar{v} \left(1-\frac{\bar{u}\bar{v}}{4}\right)^{-1} \right)\ ,
$$
we obtain
\be
\frac{2i}{
(1+v\bar{u}/4)^{2}({\bf x}_0\cdot{\bf w})^{2}}\,
\delta_{H_{d-1}}({\bf w},{\bf \bar{w}})\ .
\label{RHS}
\ee
Next we consider the LHS of (\ref{propeq}). Explicitly parametrizing the metric $d{\bf w}^2$ on $H_{d-1}$ in (\ref{AdSMetric}) as
$$
d{\bf w}^2 = \frac{d{\chi}^2}{{\chi}^2-1}+({\chi}^2-1)ds^2(S_{d-2}),
$$
where $\chi=-\mathbf{x}_0\cdot \mathbf{w}$, we have that
\bes
\Box_{\rm AdS} \Pi_\Delta=\left[\Box_{{\rm H}_{d-1}}+2\,\frac{{\chi}^2-1}{{\chi}}\partial_{\chi} \right]\Pi_\Delta + \partial_u(\cdots)\, ,
\ees
where we do not show the explicit terms of the form $\partial_u(\cdots)$ since they will vanish once integrated along the two geodesics. Integrating in $du d\bar{v}$ we conclude that (\ref{RHS}) must be equated to
$$
-\frac{2i}{(1+v\bar{u}/4)^{2}} \left[\Box_{{\rm H}_{d-1}}-\Delta(\Delta-d)
+2\,\frac{{\chi}^2-1}{{\chi}}\partial_{\chi}
\right]
\frac{G({\bf w},{\bf \bar{w}})}{{\chi}\bar{{\chi}}}\ .
$$
Using the fact that
$$
\left[\Box_{{\rm H}_{d-1}},{\chi}^{-1}\right]=\frac{1}{{\chi}}\left(-2\frac{{\chi}^2-1}{{\chi}}\partial_{\chi}+(3-d)-\frac{2}{{\chi}^2}\right)\ ,
$$
we finally deduce that
$$
\left[ \Box_{H_{d-1}} +1-d -\Delta( \Delta-d ) \right] G(  {\bf w} \cdot
{\bf \bar{w}} )
= -\delta(  {\bf w}, {\bf \bar{w}} )\ .
$$
In appendix \ref{app1} we show that this last equation is also valid for general spin $j$. We conclude that
the function $G$ is the scalar Euclidean propagator in the hyperboloid
$H_{d-1}$
of mass squared  $(\Delta-1)(\Delta-1-d+2)$ and corresponding dimension $\Delta-1$. Denoting this propagator by
$\Pi_{\perp}( {\bf w}, {\bf \bar{w}} )$,
 the eikonal amplitude can be written as
\begin{equation}
\begin{array}{rcl}
A &\simeq&
\displaystyle{(2\omega)^2 \int_{-\infty}^{\infty} dv d\bar{u}  \int_{H_{d-1}} d{\bf w} d{\bf \bar{w}}
 F_1(v,{\bf w}) F_3(v,{\bf w})F_2(\bar{u} , {\bf \bar{w}})F_4(\bar{u} , {\bf \bar{w}})}\\
&&
\ \ \ \ \ \ \ \ \ \ \ \ \ \ \  \ \ \ \ \ \ \ \ \ \ \ \ \ \ \ \ \ \ \ \ \ \ \ \ \ \ \ \ 
\displaystyle{\exp \left( \frac{ig^2  }{2}\,s^{j-1}\,
\Pi_{\perp}({\bf w},{\bf \bar{w}}) \right)}\ ,
\end{array}
\label{eikAdS}
\end{equation}
with $s$ given by (\ref{sAdS}).

\begin{figure}
\begin{center}
\includegraphics[height=6cm]{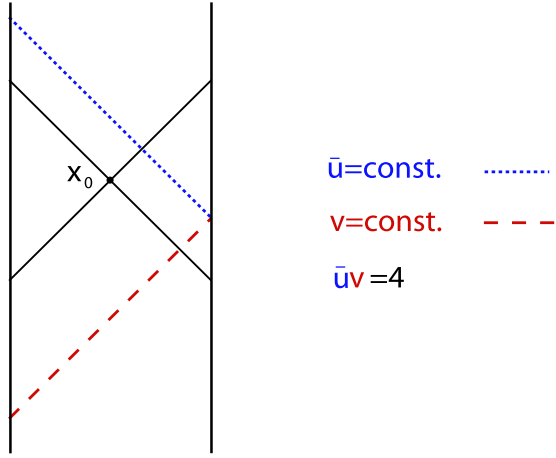}
\caption{\small The null geodesics with constant $\bar{u}=-4/v$ are the reflection   in the AdS conformal boundary
of the null geodesics with constant $v$.}
\label{reflection}
\end{center}
\end{figure}

\subsection{Localized Wave Functions}

The eikonal amplitude in AdS has a striking similarity with the
standard flat space eikonal amplitude. However, an important
difference is the factor $
\left(1+\frac{v\bar{u}}{4}\right)^{2}$ in the definition (\ref{sAdS}) of $s$, which
makes the exponent in the eikonal amplitude (\ref{eikAdS}) diverge for $v\bar{u}=-4$ and $j=0$. This divergence can be
traced back to the colinearity of the tangent vectors ${\bf k}$
and $ {\bf \bar{k}}$ of the null geodesics labeled by $\{v,{\bf
w} \}$ and $\{\bar{u} , {\bf \bar{w}} \}$ describing classical
trajectories of particle 1 and 2, respectively. When
$v\bar{u}=-4$, one null geodesic can be seen as the reflection of
the other on the AdS boundary (see Figure \ref{reflection}). Thus,
the propagator from a point on one geodesic to a point in the
other, diverges since these points are connected by a null
geodesic. This is the physical meaning of the divergence at
$v\bar{u}=-4$. Clearly, we should doubt  the accuracy of the
eikonal approximation in this case of very strong interference. To
avoid this annoying divergence, from now on we shall localize the
external wave functions of particle 1 and 2 around $v=0$ and
$\bar{u}=0$, respectively. More precisely, we shall choose
\begin{equation}
\begin{array}{rcl}
&\psi_1({\bf x})\simeq e^{-i\omega v}F(v)F_1({\bf w})\ ,\ \ \ \ \ \ \ \ \ \ \spa{.3}
&\psi_2({\bf \bar{x}})\simeq e^{-i\omega \bar{u}}F(\bar{u}) F_2( {\bf \bar{w}})\ , \\
&\psi_3({\bf x})\simeq e^{\,i\omega v} F^\star(v)F_3({\bf w})\ , \ \ \ \ \ \ \ \ \ \
&\psi_4({\bf \bar{x}})\simeq e^{\,i\omega \bar{u}} F^\star(\bar{u}) F_4( {\bf \bar{w}})\ ,
\end{array}
\label{eikwavefunc}
\end{equation}
where the profile $F(\alpha)$ is localized in the region $|\alpha|<\Lambda \ll 1$ and it
is normalized as
\begin{equation}
\int_{-\infty}^{\infty} d\alpha\,\left| F(\alpha) \right|^2 = \sqrt{2}\ .
\label{norm}
\end{equation}
On the other hand, the smoothness condition $|\partial F|\ll \omega |F|$ requires $\Lambda \gg 1/\omega$.
The two conditions,
$$
1/\omega \ll \Lambda \ll 1 \ ,
$$
are compatible for high energy scattering, when the de Broglie wavelength of the external particles
is much shorter than the radius of AdS ($\ell=1$).
With this choice of external wave functions, the amplitude simplifies to
\begin{equation}
A_{eik}
 \simeq  8\omega^2  \int_{H_{d-1}} d{\bf w} d{\bf \bar{w}}
 F_1({\bf w}) F_3({\bf w})F_2( {\bf \bar{w}})F_4({\bf \bar{w}})
\ \exp \left( \frac{ig^2}{2}\, s^{j-1}\, \Pi_{\perp}({\bf w},{\bf \bar{w}})
\right)\ ,
\label{ampsimp}
\end{equation}
where now 
\begin{equation}
s=\frac{(2\omega)^{2}}{
({\bf x}_0\cdot{\bf w})\, ({\bf x}_0\cdot  {\bf \bar{w}})}\ .
\label{sAdSlocalized}
\end{equation}

\section{Relation to the Dual CFT}\label{sectCFT}

The AdS/CFT correspondence predicts the existence of a dual
CFT$_{d}$ living on the boundary of AdS$_{d+1}$. In particular,
the AdS scattering amplitude we determined in the previous section
is directly related to the CFT four point--function of scalar
primary operators. We shall now explore this connection to find
properties of four--point functions in CFTs with AdS duals.

Firstly, we must introduce some convenient notation
\cite{Paper1,Paper2}. The boundary of AdS can be thought of as the
set of null rays through the origin of the embedding space
$\mathbb{R}^{2,d}$. More precisely, a point in the boundary of AdS
is given by
$$
{\bf p} \in \mathbb{R}^{2,d} \ ,\ \ \ \ \ \ \ \ \ \ \ \ \
{\bf p}^2=0 \ ,\ \ \ \ \ \ \ \ \ \ \ \ \
{\bf p}\sim \lambda {\bf p}\ \ \ \ (\lambda>0)\ .
$$
In this language, a CFT correlator  of scalar primary operators
located at points $\mathbf{p}_{1},\ldots,\mathbf{p}_{n}$
is described by an amplitude
\[
A\left(  \mathbf{p}_{1},\ldots,\mathbf{p}_{n}\right)
\]
invariant under $SO\left(  2,d\right)  $ and therefore only a function of the invariants
$\mathbf{p}_{i}\cdot\mathbf{p}_{j}$.
Moreover, since the boundary points $\mathbf{p}_{i}$ are defined only up to rescaling,
the amplitude $A$ will be homogeneous in each entry
\[
A\left( \ldots, \lambda\mathbf{p}_{i}, \ldots \right) =
\lambda^{-\Delta_{i}} A\left( \ldots, \mathbf{p}_{i}, \ldots \right)~,
\]
where $\Delta_{i}$ is the conformal dimension of the $i$--th scalar primary operator.

The AdS scattering amplitude considered in the previous section is directly related to the correlator
\[
A\left( \mathbf{p}_{1},\mathbf{p}_{2},\mathbf{p}_{3},\mathbf{p}_{4} \right) =
 \left\langle \mathcal{O}_{1}\left( \mathbf{p}_{1}\right)
\mathcal{O}_{2}\left( \mathbf{p}_{2}\right) \mathcal{O}_{1}\left( \mathbf{p}_{3}\right)
\mathcal{O}_{2}\left( \mathbf{p}_{4}\right) \right\rangle_{\text{\textrm{CFT}}_{d}}\ ,
\]
where the scalar primary operators $\mathcal{O}_{1}$ and $\mathcal{O}_{2}$ have dimensions
$\Delta_{1}$ and $\Delta _{2}$, respectively.
The four--point amplitude $A$ is just a function of two cross--ratios $z,\bar{z}$ which we define,
following \cite{Osborn, Osborn22}, in terms of the kinematical invariants
as  \footnote{Throughout the paper, we shall consider barred and unbarred variables as independent,
with complex conjugation denoted by $\star$. In general $\bar{z}=z^{\star}$ when considering the
analytic continuation of the CFT$_{d}$ to Euclidean signature. For Lorentzian signature, either
$\bar{z}=z^{\star}$ or both $z$ and $\bar{z}$ are real. These facts follow simply from solving
the quadratic equations for $z$ and $\bar{z}$.}
\begin{equation}
\begin{array}{rcl}
 z\bar{z}  & =&
\displaystyle{\frac{(\mathbf{p}_{1}\cdot\mathbf{p}_{3})(\mathbf{p}_{2}\cdot\mathbf{p}_{4})}
{(\mathbf{p}_{1}\cdot\mathbf{p}_{2})(\mathbf{p}_{3}\cdot\mathbf{p}_{4})}  }\ , \spa{0.6}
\\
\left(  1-z\right)  \left(  1-\bar{z}\right)   & =&
\displaystyle{\frac{(\mathbf{p}_{1}\cdot\mathbf{p}_{4})(\mathbf{p}_{2}\cdot\mathbf{p}_{3})}
{(\mathbf{p}_{1}\cdot\mathbf{p}_{2})(\mathbf{p}_{3}\cdot\mathbf{p}_{4})}}\ .
\end{array}
\label{zzbar}
\end{equation}
Then, the four--point amplitude can be written as
\[
A\left( \mathbf{p}_{1},\mathbf{p}_{2},\mathbf{p}_{3},\mathbf{p}_{4} \right)  =
 K_{\Delta_1}({\bf p}_1, {\bf p}_3 )\,
 K_{\Delta_2}({\bf p}_2, {\bf p}_4 )\,
\mathcal{A}\left(  z,\bar{z}\right)~,
\]
where $\mathcal{A}$ is a generic function of $z,\bar{z}$ and
$K_{\Delta}({\bf p}, {\bf p}')$ is the  boundary propagator  of conformal dimension $\Delta$ defined below.
With this normalization, the disconnected graph gives $ \mathcal{A}=1$.

By the AdS/CFT correspondence, CFT  correlators can be computed using
string theory in Anti de--Sitter spacetime. We shall work in the limit of small string length compared
to the radius of AdS, where the supergravity description is valid.
In this regime, the above four--point correlator  is given by the sum of
all Feynman--Witten diagrams like the one in Figure \ref{fig1}, with bulk to boundary propagators $K_{\Delta}({\bf p}, {\bf x} )$ as
external wave functions,
\begin{eqnarray*}
\psi_1({\bf x} )=K_{\Delta_1}({\bf p}_1, {\bf x} )\ ,\ \ \ \ \ \ \ \ \ \ \ \
\psi_2({\bf x} )=K_{\Delta_2}({\bf p}_2, {\bf x} )\ ,\\
\psi_3({\bf x} )=K_{\Delta_1}({\bf p}_3, {\bf x} )\ ,\ \ \ \ \ \ \ \ \ \ \ \
\psi_4({\bf x} )=K_{\Delta_2}({\bf p}_4, {\bf x} )\ .
\end{eqnarray*}
More generally, we can prepare any on--shell wave function  in the bulk by
superposing bulk to boundary propagators from many boundary points. For example,
$$
\psi_1({\bf x} )=\int_{\Sigma} d{\bf p}_1\, \phi_1({\bf p}_1 )\, K_{\Delta_1}({\bf p}_1, {\bf x} )\ ,
$$
where the boundary integration is done along a specific section $\Sigma$ of the light--cone, with metric induced
by the embedding space. Choosing a different section corresponds to conformal transformations of the
boundary.
The boundary wave function $\phi_1({\bf p}_1 )$ must be a homogeneous function of weight $\Delta_1-d$,
$$
\phi_1(\lambda {\bf p}_1 )= \lambda^{\Delta_{1}-d}\,\phi_1({\bf p}_1 )\ ,
$$
so that the integral is invariant under conformal transformations of the boundary.
Therefore, given boundary wave functions $\phi_i$, such that the corresponding
bulk  wave functions $\psi_i$ are of the eikonal type as defined in the previous section,
we have
$$
\int_{\Sigma} d{\bf p}_1\cdots  d{\bf p}_4\,
\phi_1({\bf p}_1 )\cdots \phi_4({\bf p}_4 )\,
A\left( \mathbf{p}_{1},\mathbf{p}_{2},\mathbf{p}_{3},\mathbf{p}_{4} \right)
\simeq A_{eik}\ ,
$$
where $A_{eik}$ is given by (\ref{ampsimp}).

\subsection{CFT Eikonal Kinematics}
\label{CFTeik}

In order to construct the relevant eikonal wave functions, we shall need
to analyze more carefully the global structure of AdS.
Consider a point $\mathbf{Q}$, either in AdS
or on its boundary. The future and past light--cones starting from
$\mathbf{Q}$ divide global AdS and its boundary into an infinite
sequence of regions, which we label by an integer.
Given a generic point $\mathbf{Q}^{\prime }$, we introduce the integral function 
$n\left( \mathbf{Q}^{\prime },\mathbf{Q}\right)$ which vanishes when 
$\mathbf{Q}^{\prime }$ is space--like related to $\mathbf{Q}\,$ and which increases (decreases)
as $\mathbf{Q}^{\prime }$ moves forward (backward) in global time
and crosses the light cone of $\mathbf{Q}$.
Clearly $n\left( \mathbf{Q},\mathbf{Q}^{\prime }\right) =-n\left( \mathbf{Q}^{\prime },\mathbf{Q}\right) $.
In terms of the function $n\left( \mathbf{Q}^{\prime },\mathbf{Q}\right)$, the boundary and
the bulk to boundary propagators $K_{\Delta}({\bf p}, {\bf p}' )$ and $K_{\Delta}({\bf p}, {\bf x} )$
are given by
\begin{equation}
K_{\Delta}({\bf p}, {\bf p}' )=
\frac{\mathcal{C}_{\Delta}}{|2{\bf p} \cdot {\bf p}'|^{\Delta}}\,i^{\,-2\Delta|n\left( \mathbf{p},\mathbf{p}^{\prime }\right)|}\ ,
\ \ \ \ \ \ \ \ \ \
K_{\Delta}({\bf p}, {\bf x} )=
\frac{\mathcal{C}_{\Delta}}{|2{\bf p} \cdot {\bf x}|^{\Delta}}\,i^{\,-2\Delta|n\left( \mathbf{p},\mathbf{x}\right)|}\ ,
\label{propagators}
\end{equation}
where  \footnote{The normalization $\mathcal{C}_{\Delta}$ is not the standard one used
in the literature \cite{FreedmanRev, w98}. In this paper, the boundary propagator $K_{\Delta}({\bf p}, {\bf p}')$
and the bulk to boundary propagator
$K_{\Delta}\left( \mathbf{p} , \mathbf{x}\right)$ are taken to be the limit of the bulk to bulk
propagator $\Pi_{\Delta}(\mathbf{x'},\mathbf{x})$ as the bulk points approach the boundary.
As shown in \cite{fmmr98,kw99},  naive Feynman graphs
in AdS computed with this prescription give correctly normalized CFT correlators, including
the subtle two--point function.}
$$
\mathcal{C}_{\Delta}=\frac{1}{2\pi^{\frac{d}{2}}}\frac{\Gamma\left(
\Delta\right)  }{\Gamma\left(  \Delta-\frac{d}{2}+1\right)  }\ .
$$
In particular, if $n\left( \mathbf{p},\mathbf{x}\right)=0,\pm 1$, then
$$
K_{\Delta}({\bf p}, {\bf x} )=
\frac{\mathcal{C}_{\Delta} }{( -2{\bf p} \cdot {\bf x}+i\epsilon )^{\Delta}  }\ .
$$

\begin{figure}[tbp]
\begin{center}
\includegraphics[width=9cm]{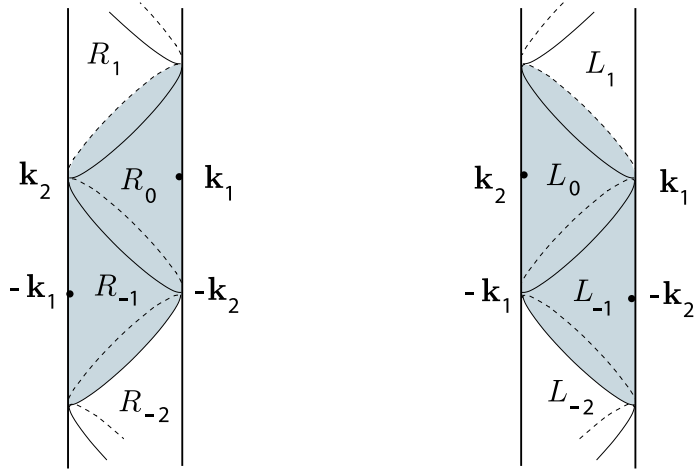}
\end{center}
\caption{{\protect\small The momenta
$\mathbf{k}_{1},\mathbf{k}_{2}$ divide AdS space in Poincar\'{e}
patches $L_n$ and $R_n$. The boundary wave functions $\phi_1$ and
$\phi_3$ ($\phi_2$ and $\phi_4$) are localized on the boundary of
$R_{-1}$ and $R_{0}$ ($L_{-1}$ and $L_{0}$). }} \label{FigPP}
\end{figure}

Recall that the momenta $\mathbf{k}_{1}$ and $\mathbf{k}_{2}$ indicate, respectively, the
outgoing directions of particles $1$ and $2$, whereas $-\mathbf{k}_{1}$ and $-\mathbf{k}_{2}$ 
indicate the incoming ones.
These null vectors are identified with boundary points as in Figure \ref{FigPP}.
We therefore expect the boundary wave functions to be localized around
these points. Implicit in the discussion
in the previous sections is the assumption that 
$n\left( \mathbf{k}_{1},\mathbf{k}_{2}\right) =n\left( -\mathbf{k}_{1},-\mathbf{k}_{2}\right) =0$,
whereas $n\left( \mathbf{k}_{2},-\mathbf{k}_{1}\right) =n\left( \mathbf{k}_{1},-\mathbf{k}_{2}\right) =1$. 
The momentum $\mathbf{k}_{1}$
divides global AdS$_{d+1}$ space into a set of Poincar\`{e}
patches $L_{n}$ of points $\mathbf{x}$ such that 
$n\left(\mathbf{x,k}_{1}\right) =n$, which are separated by the surface
$\mathbf{x}\cdot \mathbf{k}_{1}=0$, as shown explicitly in Figure
\ref{FigPP}.
Similarly, we have the patches $R_{n}$ of points $\mathbf{x}$ with
$n\left( \mathbf{x,k}_{2}\right) =n$, separated by the surface
$\mathbf{x}\cdot \mathbf{k}_{2}=0$. A point $\mathbf{x}$, either
in AdS or on its boundary, with $\mathbf{x\cdot k}_{1}<0$ ($\mathbf{x\cdot k}_{1}>0$) 
will be within a region $L_{n}$ with $n$ even (odd), and
similarly
for the regions $R_{n}$. From our previous construction, 
we see that the interaction takes
place around the hyperboloid $H_{d-1}$ defined by the intersection of the boundary between $R_{0}$ and $R_{-1}$ 
($\mathbf{x\cdot k}_{2}=0 $)
and the boundary between $L_{0}$ and $L_{-1}$ ($\mathbf{x\cdot k}_{1}=0 $). Let us then consider the
incoming wave $\phi _{1}\left( \mathbf{p}_{1}\right) $. In order to achieve
the required eikonal kinematics, we shall localize $\phi _{1}$ on the boundary
of $R_{-1}$, around the point $-\mathbf{k}_{1}$.
We shall show in the next section that, if we choose only
positive frequency modes with respect to the action of time translation in
this patch, which is generated by $\mathbf{T}_{2}$, the corresponding bulk
wave function $\psi _{1}$ will have support only on patches $R_{n}$ with $n\geq -1$. 
Similarly, we shall localize $\phi _{3}$ on $\partial R_{0}$,
around the point $\mathbf{k}_{1}$,
with negative frequency modes only, so that $\psi _{3}$
will have support on $R_{n}$ for $n\leq 0$. The overlap of $\psi_{1}$ and 
$\psi _{3}$ will then be non vanishing only in regions
$R_{-1}$ and $R_{0}$, which are those parametrized explicitly
by the coordinates $\{u,v,\mathbf{w}\}$. In a symmetric way, we shall localize 
$\phi_{2}$  ($\phi_{4}$) on $\partial L_{-1}$  ($\partial L_{0}$), around the
point  $-\mathbf{k}_{2}$ ($\mathbf{k}_{2}$), with positive
(negative) frequency modes with respect to $\mathbf{T}_{1}$.  The overlap 
of $\psi _{2}$  and $\psi _{4}$ is then localized
in regions $L_{-1}$ and $L_{0}$, parametrized by $\{\bar{u},\bar{v},\mathbf{\bar{w}}\}$. 
Summarizing, the relevant choice of kinematics for the four
points $\mathbf{p}_{i}$ ($i=1,\cdots ,4$) is given by
\begin{equation}
\begin{array}{lllll}
\mathbf{p}_{1} \sim -\mathbf{k}_{1}&\Rightarrow\ &
\mathbf{p}_{1} \,\in\, \partial R_{-1}\ \ \ \ \ \ \ \ \ \ \ \ \ 
&\left( \mathbf{p}_{1}\cdot \mathbf{k}_{2}>0\right) \ ,  \spa{0.2}\\
\mathbf{p}_{2} \sim -\mathbf{k}_{2}&\Rightarrow&
\mathbf{p}_{3} \,\in\, \partial R_{0}
&\left( \mathbf{p}_{3}\cdot \mathbf{k}_{2}<0\right) \ , \spa{0.2} \\
\mathbf{p}_{3} \sim \mathbf{k}_{1}&\Rightarrow&
\mathbf{p}_{2} \,\in\, \partial L_{-1}
&\left( \mathbf{p}_{2}\cdot \mathbf{k}_{1}>0\right) \ ,  \spa{0.2}\\
\mathbf{p}_{4} \sim \mathbf{k}_{2}&\Rightarrow&
\mathbf{p}_{4} \,\in\, \partial L_{0}
&\left( \mathbf{p}_{4}\cdot \mathbf{k}_{1}<0\right) \ ,
\end{array}
\label{boundp}
\end{equation}
so that
\begin{equation*}
\begin{array}{l}
n\left( \mathbf{p}_{1},\mathbf{p}_{2}\right)  = 
n\left( \mathbf{p}_{3},\mathbf{p}_{4}\right) =0~,   \spa{0.2} \\
n\left( \mathbf{p}_{4},\mathbf{p}_{1}\right)  =
n\left( \mathbf{p}_{3},\mathbf{p}_{2}\right) =1~.  
\end{array}
\end{equation*}
We shall choose, once and for all, a
specific normalization of the $\mathbf{p}_{i}$ by rescaling the external
points, so that
\[
2\mathbf{p}_{1}\cdot \mathbf{k}_{2}=-2\mathbf{p}_{3}\cdot \mathbf{k}_{2}=
2\mathbf{p}_{2}\cdot \mathbf{k}_{1}=-2\mathbf{p}_{4}\cdot \mathbf{k}_{1}=\left( 2\omega \right) ^{2}~.
\]
It is also convenient to parametrize the $\mathbf{p}_{i}$ in terms of Poincar\'{e}
coordinates. Using the explicit coordinates on $\mathbb{R}^{2,d}\simeq
\mathbb{M}^{2}\times \mathbb{M}^{d}$ introduced in section \ref{NCWF}, we
write
\begin{equation}
\begin{array}{ll}
\displaystyle{\mathbf{p}_{1} = 2\omega \Big( p_{1}^2 \,,1\,,p_{1}\Big)} ~,\ \ \ \ \ \ \ \ \ \ \ \ \ \ \ \ \ \ \ \ 
&\displaystyle{\mathbf{p}_{2}= 2\omega \Big( 1,\,p_{2}^2\,,p_{2}\Big)} ~,  \spa{0.4}\\
\displaystyle{\mathbf{p}_{3} =-2\omega \Big( p_{3}^2\,,1\,,p_{3}\Big)}~,\ \ \ \ \ \ \ \ \ \ \ \ \ \ \ \ \ \ 
&\displaystyle{\mathbf{p}_{4}=-2\omega \Big( 1\,,p_{4}^2\,,p_{4}\Big)} ~,
\end{array}
\label{pcoord}
\end{equation}
with the Poincar\'{e} positions $p_{i}\in \mathbb{M}^{d}$ small, i.e. in components $| p_{i}^{\,a}|\ll 1$.

\begin{figure}[tbp]
\begin{center}
\includegraphics[width=5.5cm]{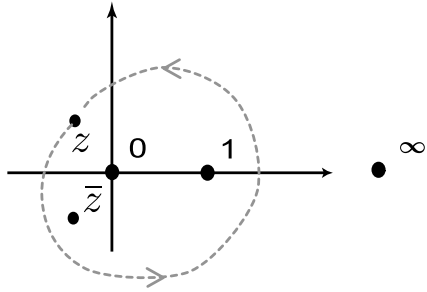}
\end{center}
\caption{{\protect\small Analytic continuation necessary to obtain
$\hat{\mathcal{A}}$ from the Euclidean amplitude ${\mathcal{A}}$.}}
\label{FigCC}
\end{figure}

We shall denote the corresponding CFT amplitude, computed with this
kinematics, by
\begin{equation}
\hat{A}\left( \mathbf{p}_{1},\cdots ,\mathbf{p}_{4}\right) =K_{\Delta
_{1}}\left( \mathbf{p}_{1},\mathbf{p}_{3}\right) K_{\Delta _{2}}\left(
\mathbf{p}_{2},\mathbf{p}_{4}\right)  \hat{\mathcal{A}}\left( z,\bar{z}
\right) ~,  
\label{zzz1}
\end{equation}
where the cross ratios are small and satisfy
\be
z\bar{z} \simeq p^{2}\bar{p}^{2}~,~\ \ \ \ \ \ \ \ \ \ \ \ \ \ \ \ \ \ \ \ 
z+\bar{z} \simeq 2p \cdot \bar{p}~,
\label{cba}
\ee
with $$p=p_3-p_1\, ,\ \ \ \ \ \ \ \ \ \ \ \ \ \ \ \ \ \bar{p}=p_2-p_4\ .$$
We shall reserve the label $A$ and $\mathcal{A}$ for the amplitude computed
on the principal Euclidean sheet, where $n\left( \mathbf{p}_{i},\mathbf{p}_{j}\right) =0$. 
As we shall discuss in detail in section \ref{sectAC}, the
amplitude $\hat{\mathcal{A}}\left( z,\bar{z}\right) $ is related to 
$\mathcal{A}\left( z,\bar{z}\right) $ by analytic continuation. More
precisely, we shall show that
\be
\hat{\mathcal{A}}\left( z,\bar{z}\right) =\mathcal{A}^{\circlearrowleft
}\left( z,\bar{z}\right) ~,
\label{CCeq}
\ee
where the right--hand side indicates the function obtained by keeping $\bar{z}$ 
fixed and rotating $z$ counter--clockwise  around the branch points  $0$ and $1$, 
as shown in Figure \ref{FigCC}.

Let us now discuss the boundary propagators $K_{\Delta }$ in (\ref{zzz1}).
The only subtle issue comes from the appropriate phase factors \cite{Paper1}. 
More precisely, given the choices in (\ref{boundp}) and the form of the boundary
propagator in (\ref{propagators}), 
we have that $K_{\Delta _{1}}\left( \mathbf{p}_{1},\mathbf{%
p}_{3}\right) $ is given by $\mathcal{C}_{\Delta _{1}}\left\vert 2\mathbf{p}%
_{1}\cdot \mathbf{p}_{3}\right\vert ^{-\Delta _{1}}$ times the following
phases
\begin{eqnarray}
&&1~\ \ \ \ \ \ \ \ \ \ \ \ \ \ \ \ \ \ \ \ 
\mathbf{p}_{1},\mathbf{p}_{3}\text{ spacelike separated}  \nonumber \\
&&i^{-2\Delta _{1}}~\ \ \ \ \ \ \ \ \ \ \ \ \ \ \ 
\mathbf{p}_{3}\text{ in the future of }\mathbf{p}_{1}\text{ with }\mathbf{p}_{1}\cdot \mathbf{p}_{3}>0  
\label{phases} \\
&&i^{-4\Delta _{1}}~\ \ \ \ \ \ \ \ \ \ \ \ \ \ \ 
\mathbf{p}_{3}\text{ in the future of }\mathbf{p}_{1}\text{ with }\mathbf{p}_{1}\cdot \mathbf{p}_{3}<0  \nonumber
\end{eqnarray}
A similar statement applies to the propagator $K_{\Delta _{2}}\left( \mathbf{%
p}_{2},\mathbf{p}_{4}\right) $.
The amplitude (\ref{zzz1}) is then given, in terms of $p ,\bar{p}$ by
\begin{equation}
\hat{A}\left( p ,\bar{p}\right) = \frac{(2\omega\,i)^{-2\Delta _{1}} \mathcal{C}%
_{\Delta _{1}}}{\left( p ^{2}+i\epsilon _{p }\right) ^{\Delta _{1}}}\,
\frac{(2\omega\,i)^{-2\Delta _{2}}\mathcal{C}_{\Delta _{2}}}{\left( \bar{p}%
^{2}-i\epsilon _{\bar{p}}\right) ^{\Delta _{2}}} ~\hat{\mathcal{A}}
\left( z,\bar{z}\right)\ ,   \label{zzz2}
\end{equation}
where we have explicitly written the two propagators using
\[
\epsilon _{p }=\epsilon \func{sign}\left( -x_0\cdot p \right) ~,~
\]
which picks the correct branch of the logarithm consistent with the phase
prescription in (\ref{phases}). Notice that $x_0$ is any future directed vector in $\mathbb{M}^d$,
which we choose to be the reference point introduced in (\ref{x0}) of section  \ref{eikonalAdS}.

\subsection{Boundary Wave Functions}\label{boundfunc}

\begin{figure}
\begin{center}
\includegraphics[height=4.8cm]{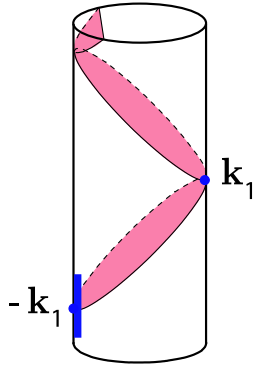}
\caption{\small
The boundary wave function  $\phi_1$ is localized along a small timelike segment centered in   $-{\bf k}_1$.
The bulk wave function $\psi_1$ is mainly supported around the region ${\bf k}_1\cdot {\bf x}=0 $
in the future of the boundary point $-{\bf k}_1$.}
\label{psi1}
\end{center}
\end{figure}

We shall now describe in detail a particularly convenient choice of boundary wave functions, consistent with the general description
of the previous section, and which correspond to bulk wave functions of the eikonal type.
First recall that in section \ref{NCWF}, ${\bf k}_1$ defined a surface in AdS containing 
the null geodesics that go from the boundary point $-{\bf k}_1$ to  ${\bf k}_1$.
We have then used the AdS isometry generated by $ {\bf T}_2$ to build the congruence of null
geodesics associated to particle 1. 
This isometry is time translation in the Poincar\'e patch $R_{-1}$, with boundary centered
at  $-{\bf k}_1$.
It is then natural
to localize the boundary wave function of $\mathcal{O}_{1}$ along the timelike line
$$
{\bf p}_1(t)=-e^{\,t {\bf T}_2}\,{\bf k}_1=-{\bf k}_1+t\omega \,{\bf x}_0+\frac{t^2}{4}\,{\bf k}_2\ .
$$
In fact, parametrizing ${\bf p}_1(t)$ in Poincar\'e coordinates as in (\ref{pcoord}), we have that
$$
p_1(t) = \frac{t}{2}\, x_0 \ ,
$$
so, as a function of $t$, we are moving in the future time direction indicated by $x_0$.
We then modulate the boundary function with $\omega\,F(t)\, e^{-i\omega t} $,
where the function $F$ is the profile function introduced in (\ref{eikwavefunc}).
The bulk wave function $\psi_1$ is then given by
$$
\psi_1({\bf x} )=\omega
\int dt \,F(t)\, e^{-i\omega t}\,\frac{\mathcal{C}_{\Delta_1} }{\big( -2{\bf p}_1(t) \cdot {\bf x}+i\epsilon \big)^{\Delta_1}}\ ,
$$
where the $i\epsilon$ prescription is correct for all points ${\bf x}$ in region $R_{-1}$. 
Since $F(t)$ is non--vanishing only for $|t|< \Lambda$, the above description is valid 
also in part of region $R_{0}$, as we shall show shortly.
In the coordinate system (\ref{coordAdS}), valid in $R_{-1}$ and $R_{0}$,  we have
$$
-2{\bf p}_1(t) \cdot {\bf x} = -2\omega (t-v)\left(1+\frac{u}{4} (t-v) \right)({\bf x}_0 \cdot {\bf w} ) \ ,
$$
showing that the integrand diverges for $t=v$ and $t=v-4/u$. The first divergence corresponds 
to the future directed signal from point
${\bf p}_1(t)$, whereas the second divergence comes from the reflection at the AdS boundary
 for $u>0$ and from the backward signal from ${\bf p}_1(t)$ for $u<0$. The backwards signal is 
relevant in region $R_{-1}$, where the $i\epsilon$ prescription is valid. For positive $\omega$
 one may close the $t$ contour avoiding completely the singularity from the backwards signal,
 showing that positive frequencies propagate forward in global time. In region $R_0$, 
on the other hand, the $i\epsilon$ prescription is valid up to the reflected signal at
 the second singularity, more precisely for $u\Lambda<|4-vu|$. In this part of $R_0$ and in region
 $R_{-1}$, for large $\omega$, the integral is dominated by the divergence at $t=v$, and we have that
$$
\psi_1({\bf x} ) \simeq \omega
\,F(v)\, \int dt \, e^{-i\omega t}\,\frac{ \mathcal{C}_{\Delta_1} }{\big(-2\omega (t-v)({\bf x}_0 \cdot {\bf w} )+i\epsilon \big)^{\Delta_1}}\ .
$$
It is then clear that, for large $\omega$, the wave function $\psi_1$ has precisely the required form
$$
\psi_1({\bf x} )\simeq e^{-i\omega v}\,F(v) F_1( {\bf w} )\ ,
$$
with
$$
F_1( {\bf w} )= i^{-\Delta_1} \,\frac{2\pi\,\mathcal{C}_{\Delta_1}}{\Gamma(\Delta_1)}
 \left( -2{\bf x}_0 \cdot {\bf w} \right)^{-\Delta_1}  \ .
$$
Thus, the wave function  $\psi_1$ is supported mainly around the future directed null geodesics starting from
the point $-{\bf k}_1$ of the boundary, as depicted in Figure \ref{psi1}.
Similarly, we choose the boundary wave function of $\mathcal{O}_{2}$
localized along the timelike line
$$
{\bf p}_2(t)=-e^{\,t {\bf T}_1}\,{\bf k}_2=-{\bf k}_2+t\omega \,{\bf x}_0+\frac{t^2}{4}\,{\bf k}_1\ ,
$$
which means 
$$
p_2(t) = \frac{t}{2}\,x_0\ .
$$
The bulk wave function  $\psi_2$ has then the required eikonal form in (\ref{eikwavefunc}) with
$$
F_2( {\bf \bar{w}} )=i^{-\Delta_2} \,\frac{2\pi\,\mathcal{C}_{\Delta_2} }{\Gamma(\Delta_2)}
 \left( -2{\bf x}_0 \cdot {\bf \bar{w}} \right)^{-\Delta_2} \ .
$$

The boundary wave functions $\phi_3$ and $\phi_4$ will be the complex conjugates of $\phi_1$ and $\phi_2$,
but localized along slightly different curves,
$$
{\bf p}_3(t)=e^{\,t {\bf T}_2}\,( {\bf k}_1 + {\bf q}) \ ,\ \ \ \ \ \ \ \ \ \ \ \ \ \ \ \ \
{\bf p}_4(t)=e^{\,t {\bf T}_1}\,( {\bf k}_2 + {\bf \bar{q}}) \ .
$$
In analogy with flat space, the eikonal regime corresponds to ${\bf q}^2, {\bf \bar{q}}^2 \ll \omega^2 $.
The fact that ${\bf p}_3 $ and  ${\bf p}_4$ must be null vectors yields the conditions
$$
{\bf q}^2=- 2{\bf k}_1 \cdot {\bf q}\ ,\ \ \ \ \ \ \ \ \ \ \ \ \ \ \ \ \
{\bf \bar{q}}^2=-2  {\bf k}_2 \cdot {\bf \bar{q}}\ .
$$
The parts of $ {\bf q} $ and ${\bf \bar{q}}$ that are,
respectively,  proportional to ${\bf k}_1 $ and ${\bf k}_2 $ are
irrelevant since we stay in the same null rays. This freedom can
be used to fix
$$
{\bf k}_2 \cdot {\bf q}=0\ ,\ \ \ \ \ \ \ \ \ \ \ \ \ \ \ \ \
{\bf k}_1 \cdot {\bf \bar{q}}=0\ .
$$
Furthermore, we shall choose ${\bf q}$  and ${\bf \bar{q}}$ orthogonal to ${\bf x}_0$.
In the explicit coordinates for $\mathbb{M}^2 \times\mathbb{M}^d $ we have
$$
{\bf q} =  2\omega \Big( q^2 ,0,q\Big)\, ,\ \ \ \ \ \ \ \ \ \ \ \ \ 
{\bf \bar q} = 2\omega \Big(0,\bar{q}^2,\bar{q}\Big)\, ,
$$
with $q\cdot x_0 = \bar{q}\cdot x_0=0$, so that
$$
p_3(t) = \frac{t}{2}\,x_0-q\, ,\ \ \ \ \ \ \ \ \ \ \ \ \ p_4(t) = \frac{t}{2}\,x_0-\bar{q}\, .
$$
We then have that
\begin{eqnarray*}
&&{\bf p}_3(t)\cdot {\bf x} = - {\bf p}_1(t)\cdot {\bf x} + {\bf q}\cdot {\bf w}
+\frac{u }{4\omega}\, ({\bf x}_0 \cdot {\bf w} )\,{\bf q}^2
\ ,\\
&&{\bf p}_4(t)\cdot {\bf \bar{x}} =  -{\bf p}_2(t)\cdot {\bf  \bar{x}} + {\bf  \bar{q}}\cdot {\bf  \bar{w}}
+\frac{\bar{v}}{4\omega} \, ( {\bf x}_0 \cdot {\bf \bar{w}}  )\,{\bf \bar{q}}^2
 \ .
\end{eqnarray*}
At large $\omega$, the leading contribution to the bulk wave function $\psi_3$ is given by
\begin{eqnarray*}
\psi_3({\bf x} )
&=&
\omega\int dt \,F^\star(t)\, e^{i\omega t}\,
\frac{\mathcal{C}_{\Delta_1}}{\big( -2{\bf p}_3(t) \cdot {\bf x}+i\epsilon \big)^{\Delta_1}}\\
&\simeq&
e^{i\omega v}\,F^\star(v) F_3( {\bf w} )\ ,
\end{eqnarray*}
where the transverse modulation function  $ F_3( {\bf w} )$ is
\begin{eqnarray*}
F_3( {\bf w} )&=&
\mathcal{C}_{\Delta_1}
\int dl \, e^{il} \big(2({\bf x}_0 \cdot {\bf w})   l - 2{\bf q}\cdot {\bf w}  +i\epsilon \big)^{-\Delta_1}  \\
&=& i^{-\Delta_1}\,\frac{2\pi\,\mathcal{C}_{\Delta_1}}{\Gamma(\Delta_1)}
 \left( -2{\bf x}_0 \cdot {\bf w} \right)^{-\Delta_1}
\exp\left(i\frac{{\bf q}\cdot {\bf w} }{{\bf x}_0 \cdot {\bf w} }\right)\ .
\end{eqnarray*}
Similarly, $\psi_4$ has the form in (\ref{eikwavefunc}) with
\begin{equation*}
F_4( {\bf \bar{w}} )
=i^{-\Delta_2} \,\frac{2\pi\,\mathcal{C}_{\Delta_2}}{\Gamma(\Delta_2)}
 \left( -2{\bf x}_0 \cdot {\bf \bar{w}} \right)^{-\Delta_2}
\exp\left(i\frac{{\bf \bar{q}}\cdot {\bf \bar{w}} }{{\bf x}_0 \cdot {\bf \bar{w}} }\right)\ .
\end{equation*}

With the specific choice of wave functions just described, the AdS eikonal
amplitude (\ref{ampsimp}) becomes
\begin{equation}
\begin{array}{r}
\displaystyle{A_{eik} \simeq 2i^{-2\Delta _{1}}i^{-2\Delta _{2}}
\left( \frac{8\pi^{2}\omega \,\mathcal{C}_{\Delta _{1}}\mathcal{C}_{\Delta _{2}}}{\Gamma
(\Delta _{1})\Gamma (\Delta _{2})}\right) ^{2}\int_{H_{d-1}}d\mathbf{w}d
\mathbf{\bar{w}}\left( -2\mathbf{x}_{0}\cdot \mathbf{w}\right) ^{-2\Delta
_{1}}\left( -2\mathbf{x}_{0}\cdot \mathbf{\bar{w}}\right) ^{-2\Delta _{2}}}
\\ \\
\displaystyle{
\exp \left( i\frac{\mathbf{q}\cdot \mathbf{w}}{\mathbf{x}_{0}\cdot 
\mathbf{w}}+i\frac{\mathbf{\bar{q}}\cdot \mathbf{\bar{w}}}{\mathbf{x}_{0}\cdot
\mathbf{\bar{w}}}+\frac{ig^{2}}{2}\left( 2\omega \right) ^{2j-2}\frac{\Pi
_{\perp }(\mathbf{w},\mathbf{\bar{w}})}{\left( (\mathbf{x}_{0}\cdot \mathbf{w})
(\mathbf{x}_{0}\cdot \mathbf{\bar{w}})\right) ^{j-1}}\right)} \ .
\end{array}
\label{eikEQ}
\end{equation}
By construction, the above expression should approximate, in the limit of
large $\omega $, the CFT correlator $\hat{A}\left( p,\bar{p} \right) $ in (\ref{zzz2})
integrated against the corresponding boundary
wave--functions $\phi _{i}\left( \mathbf{p}_{i}\right) $, 
\begin{equation}
A_{eik} \simeq\omega ^{4}\int dt_{1}\cdots dt_{4}\, F( t_{1}) F(t_{2}) F^{\star}(t_{3}) F^{\star}(t_{4})
\,e^{i\omega \left( t_{3}-t_{1}\right) +i\omega \left( t_{4}-t_{2}\right) }
\hat{A}\big( p(t_i),\bar{p}(t_i) \big) \,,  \label{CFTamp}
\end{equation}
with 
$$
p(t_i)=\frac{t_3-t_1}{2}\,x_0 - q\ ,\ \ \ \ \ \ \ \ \ \ \ 
\bar{p}(t_i)=\frac{t_2-t_4}{2}\,x_0 +\bar{q}\ .
$$
Before deriving the consequences of this result, we must clarify the structure
of the four point correlator $\hat{A}$ in (\ref{CFTamp}).
We shall devote the next three sections to this purpose and return to  equations (\ref{eikEQ}) and (\ref{CFTamp})
only in section \ref{anomdim}.


\subsection{Analytic Continuation}

\label{sectAC}

Let us discuss the issue of analytic continuation of
the amplitude $A\left( \mathbf{p}_{i}\right) $, showing in
particular how to derive (\ref{CCeq}). First note that the cross
ratios $z,\bar{z}$ as defined in (\ref{zzbar}) are invariant under
rescalings $\mathbf{p}_{i}\rightarrow \lambda _{i}\mathbf{p}_{i}$,
with $\lambda _{i}$ arbitrary and, in particular, negative.
Moreover, two different boundary points differing by a $2\pi $
translation in AdS global time have the same embedding
representation and therefore also give rise to the same values of
$z,\bar{z}$. On
the other hand, in global AdS, different sets of boundary points $\mathbf{p}%
_{i}$ with the same values of $z,\bar{z}$ have, in general, different
reduced amplitudes $\mathcal{A}\left( z,\bar{z}\right) $ related by analytic
continuation. More precisely, the amplitude $\mathcal{A}$ is a multi--valued
function of $z,\bar{z}$ with branch points at $z,\bar{z}=0,1,\infty $, and
different sets $\{\mathbf{p}_{i}\}$ with the same cross ratios correspond,
in general, to different sheets. The best way to understand this is to start
from the Euclidean reduced four--point amplitude $\mathcal{A}\left( z,\bar{z}%
\right) $ and then Wick rotate to the Lorentzian setting.

\begin{figure}[tbp]
\begin{center}
\includegraphics[width=16cm]{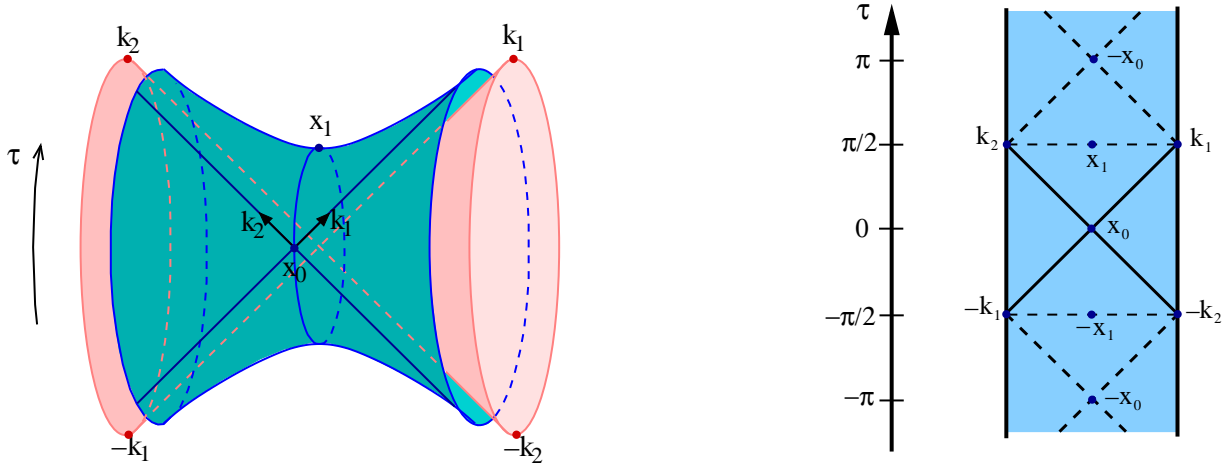}
\end{center}
\caption{{\protect\small Unwrapping the AdS$_2$ global time circle.}}
\label{globaltime}
\end{figure}

We start by choosing a global time $\tau $ in AdS. From the
embedding space perspective, global time translations are
rotations in a timelike plane. We choose this to be
the plane generated by the normalized timelike vectors $\mathbf{x}_{0}$ 
and $\mathbf{x}_{1}$, with $2\omega \,\mathbf{x}_{1}=\mathbf{k}_{1}+\mathbf{k}_{2}$ 
(see Figure \ref{globaltime}). A generic boundary point $\mathbf{p}$ can then be written as
\[
\mathbf{p}=\lambda \big[\cos (\tau )\,\mathbf{x}_{0}+\sin (\tau )\,\mathbf{x}_{1}+\mathbf{n}\big]\ ,
\]
where the vector $\mathbf{n}$ belongs to the $(d-1)$--dimensional
unit sphere embedded in the space $\mathbb{R}^{d}$ orthogonal to
$\mathbf{x}_{0}$ and $\mathbf{x}_{1}$, and the constant $\lambda>0$ depends on the choice of 
representative $\mathbf{p}$ for each
null ray. We can then consider, for
each of the boundary points under consideration, the standard Wick rotation 
$\tau \rightarrow -i\tau $ parametrized by $0\leq \theta \leq 1$,
\[
\mathbf{p}=\lambda \left[ \cos \left( -i\tau e^{\frac{i\pi }{2}\theta}\right) 
\,\mathbf{x}_{0}+\sin \left( -i\tau e^{\frac{i\pi }{2}\theta}\right) 
\,\mathbf{x}_{1}+\mathbf{n}\right] \ ,
\]
where $\theta =0$ corresponds to the Euclidean setting and $\theta
=1$ to the Minkowski one. Given the coordinates $\tau _{i}$
and $\mathbf{n}_{i}$ of the four boundary points
$\mathbf{p}_{i}$, the corresponding variables
$z ( \theta )  $, $\bar{z} ( \theta )  $ define
two paths in the complex plane parametrized by $0\leq \theta \leq
1$. The paths $z ( \theta )  $, $\bar{z} ( \theta
) $ are explicitly obtained by replacing
\[
\mathbf{p}_{i}\cdot \mathbf{p}_{j}\rightarrow \mathbf{n}_{i}\cdot \mathbf{n}%
_{j}-\cos \left( -i(\tau _{i}-\tau _{j})e^{\frac{i\pi }{2}\theta }\right) \ ,
\]
in the expressions (\ref{zzbar}). 
The Lorentzian amplitude $\hat{\mathcal{A}}$ is then given by the analytic
continuation of the basic Euclidean amplitude $\mathcal{A}$ following the
paths $z( \theta ) $, $\bar{z}( \theta )$ from $\theta =0$ to $\theta =1$.

\begin{figure}[tbp]
\begin{center}
\includegraphics[width=16cm]{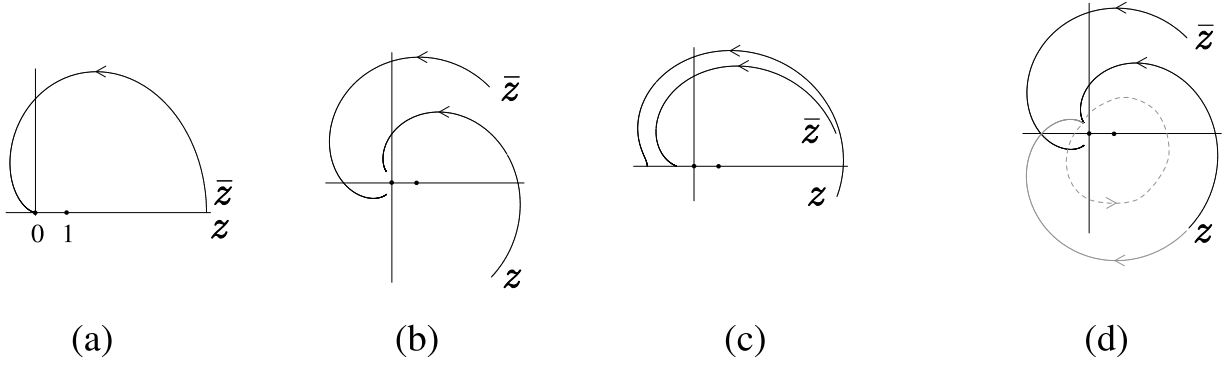}
\end{center}
\caption{{\protect\small Figures (a), (b) and (c) show the curves $z(\theta)$ and ${\bar z}(\theta)$
starting from the Euclidean setting at $\theta=0$, with $z(0)={\bar z}^{\star}(0)$.
Plot (a) corresponds to the limiting path 
$z(\protect\theta) =\bar{z}( \protect\theta)$
where $t_{i}=0$ and $\mathbf{q}=\mathbf{\bar{q}}=0$. Plots (b) and (c) correspond to general paths.
Figure (d) shows the relevant
analytic continuation relating $\hat{\mathcal{A}}$ to $\mathcal{A}$. Starting from path (b),
the curve $z(\theta)$, shown in black, is equivalent to the path shown in gray, which, in turn, is composed
of two parts. The continuous part, which is the complex conjugate of the curve $\bar z (\theta)$, computes
$\mathcal{A}$ on the principal sheet. The dashed part, also shown in Figure \ref{FigCC}, 
rotates $z$ counter--clockwise around the singularities at $0$ and $1$.
Therefore $\hat{\mathcal{A}}=\mathcal{A}^{\circlearrowleft}$.}}
\label{complexpaths}
\end{figure}

In our particular case, we have
\begin{eqnarray*}
\tau _{1}\simeq -\frac{\pi }{2}+t_{1}\ ,&\ \ \ \ \ \ \ \ &\mathbf{n}%
_{1}\simeq \frac{1}{2\omega }\left( \mathbf{k}_{2}-\mathbf{k}_{1}\right) \ ,
\\
\tau _{2}\simeq -\frac{\pi }{2}+t_{2}\ ,&\ \ \ \ \ \ \ \ &\mathbf{n}%
_{2}\simeq \frac{1}{2\omega }\left( \mathbf{k}_{1}-\mathbf{k}_{2}\right) \ ,
\\
\tau _{3}\simeq \frac{\pi }{2}+t_{3}\ ,\ \ &\ \ \ \ \ \ \ \ \ \ &\mathbf{n}%
_{3}\simeq \frac{1}{2\omega }\left( \mathbf{k}_{1}-\mathbf{k}_{2}+2\mathbf{q}%
+\frac{\mathbf{q}^{2}}{2\omega ^{2}}\,\mathbf{k}_{2}\right) \ , \\
\tau _{4}\simeq \frac{\pi }{2}+t_{4}\ ,\ \ &\ \ \ \ \ \ \ \ \ \ &\mathbf{n}%
_{4}\simeq \frac{1}{2\omega }\left( \mathbf{k}_{2}-\mathbf{k}_{1}+2\mathbf{%
\bar{q}}+\frac{\mathbf{\bar{q}}^{2}}{2\omega ^{2}}\,\mathbf{k}_{1}\right) \ ,
\end{eqnarray*}%
in the relevant regime of $t_{i}\ll 1$ and $\mathbf{q}^{2},\mathbf{\bar{q}}^{2}\ll \omega ^{2}$. 
Therefore, the complex paths $z( \theta ) ,\bar{z}( \theta )$ will be small deformations of the paths
\[
z( \theta) =\bar{z}( \theta) =\cos ^{2}\left( ie^{\frac{i\pi }{2}\theta }\pi /2\right)
\]
obtained in the special case $t_{i}=0$ and $\mathbf{q}=\mathbf{\bar{q}}=0$.
This limiting path is plotted in Figure \ref{complexpaths}a. We also show,
in Figures \ref{complexpaths}b and \ref{complexpaths}c 
two generic paths, respectively with Lorentzian values $\bar{z}(1) =z^{\star}(1)$ and 
$\func{Im}z(1)=\func{Im}\bar{z}( 1) =0$. The equations
governing the generic paths are rather cumbersome and are not
important for our present purpose.
At this point we notice that the paths $z(\theta)$ in  Figures \ref{complexpaths}b and \ref{complexpaths}c can be
continuously deformed, without crossing any branch point, to the path complex
conjugate to $\bar{z}(\theta)$, plus a full counter--clockwise turn around $0$ and $1$, 
as shown in Figure \ref{complexpaths}d. Thus, the Lorentzian amplitude $%
\hat{\mathcal{A}}(z,\bar{z})$ is obtained from the basic Euclidean amplitude
$\mathcal{A}(z,\bar{z})$ after transporting $z$ anti--clockwise around
$0$ and $1$ keeping $\bar{z}$ fixed,
\[
\hat{\mathcal{A}}(z,\bar{z})=\mathcal{A}^{\circlearrowleft}(z,\bar{z})\ .
\]


\subsection{Anomalous Dimensions as Phase Shift}


As explained in sections \ref{CFTeik} and \ref{boundfunc}, the AdS eikonal regime probes the Lorentzian 
amplitude $\hat{\mathcal{A}}$ for small values of the cross ratios $z,\bar{z}$. 
Here we shall relate the behavior of $\hat{\mathcal{A}}$ in this regime to the 
anomalous dimensions of the composite primary operators  \footnote{%
We will use this schematic notation to represent the primary composite
operators of spin $J$ and conformal dimension $E$, avoiding the rather
cumbersome exact expression.},
\[
\mathcal{O}_{1}\partial _{\mu _{1}}\cdots \partial _{\mu _{J}}\partial ^{2n}%
\mathcal{O}_{2}\ ,
\]
of large dimension $E=\Delta_1 + \Delta_2 +J+2n$ and large spin $J$. We shall also use the
conformal dimensions $h\geq \bar{h}\geq 0$ defined by
\[
E=h+\bar{h}\ ,\ \ \ \ \ \ \ \ \ \ J=h-\bar{h}\ .
\]

Consider the expansion of the Euclidean amplitude $\mathcal{A}$ in 
\emph{S}--channel conformal partial waves, corresponding to the 
OPE at $z,\bar{z}\rightarrow \infty $ (or $\mathbf{p}_{1}\rightarrow \mathbf{p}_{2}$).
Following  \cite{Paper2}, we shall assume  that the \emph{S}--channel 
decomposition of the Euclidean amplitude $\mathcal{A}$ at
large $h,\bar{h}$ is dominated by the $\mathcal{O}_{1}\mathcal{O}_{2}$
composites. Denoting their anomalous dimensions by $2\Gamma (h,\bar{h})$, we
can write
\begin{equation}
\mathcal{A}(z,\bar{z})\simeq \sum_{h\geq \bar{h} }\left( 1+R(h,%
\bar{h})\right) ~\mathcal{S}_{h+\Gamma (h,\bar{h}),\bar{h}+\Gamma (h,\bar{h}%
)}(z,\bar{z})\ ,  \label{generalex}
\end{equation}%
where $\mathcal{S}_{h,\bar{h}}$ are the partial waves corresponding to the
\emph{S}--channel exchange of a primary field with conformal dimensions $h,%
\bar{h}$. The coefficient $R(h,\bar{h})$ encodes the three point coupling
between $\mathcal{O}_{1}$, $\mathcal{O}_{2}$ and the exchanged composite  primary
field. The sum is over the lattice
\[
h,\bar{h}\in \frac{\Delta_1+\Delta_2}{2} +\mathbb{N}_{0}\ ,\ \ \ \ \ \ \ \ \ \ \ \ \ \ \ \ \ \ \
\frac{\Delta_1+\Delta_2}{2} \leq \bar{h}\leq h~.
\]%
In \cite{Paper2} we introduced an impact parameter representation $\mathcal{I%
}_{h,\bar{h}}$ for the \emph{S}--channel partial waves $\mathcal{S}_{h,\bar{h%
}}$, which approximates the latter for small $z,\bar{z}$. Moreover, we
showed that in the regime of small $z,\bar{z}$ one can replace the sum over
\emph{S}--channel partial waves in (\ref{generalex}) by an integral over
their impact parameter representation,
\[
\mathcal{A}(z,\bar{z})\simeq \int dh d\bar{h}\left( 1+R(h,\bar{h})\right) ~%
\mathcal{I}_{h+\Gamma (h,\bar{h}),\bar{h}+\Gamma (h,\bar{h})}(z,\bar{z})\ .
\]%
Expanding in powers of $\Gamma $ and dropping the explicit reference to $h,%
\bar{h}$, this equation reads
\begin{eqnarray*}
\mathcal{A}(z,\bar{z}) &\simeq &\int dhd\bar{h}\left( 1+R\right) \left(
1+\Gamma \partial +\frac{1}{2}\Gamma ^{2}\partial ^{2}+\frac{1}{3!}\Gamma
^{3}\partial ^{3}+\cdots \right) ~\mathcal{I}_{h,\bar{h}}(z,\bar{z}) \\
&\simeq &\int dhd\bar{h}\left[ 1-(\partial \Gamma -R)+\partial \big( \Gamma
(\partial \Gamma -R)\big) -\frac{1}{2}\partial ^{2}\left( \Gamma
^{2}(\partial \Gamma -R)\right) +\cdots \right] \mathcal{I}_{h,\bar{h}}(z,%
\bar{z})\ ,
\end{eqnarray*}%
where $\partial $ denotes $\partial _{h}+\partial _{\bar{h}}$ and in the
second equation we have integrated by parts inside the integral over
conformal weights $h,\bar{h}$. On one hand, the standard OPE guarantees that
the Euclidean amplitude $\mathcal{A}$ is regular for small values of $z,\bar{%
z}$. As shown in \cite{Paper2}, this implies that the coefficients of the above S--channel partial wave
expansion vanish for large $h,\bar{h}$. On the other hand, the coefficients $%
R$ and the anomalous dimensions $\Gamma $ are computed in perturbation
theory with a leading contribution at order $g^{2}$. Therefore, the
consecutive terms in the last expression have increasing leading order in
the coupling $g^{2}$ and can not cancel among themselves. We then conclude
that  \footnote{%
More precisely, $R-\partial \Gamma$ has to go to zero, for $h,\bar h \to \infty$,
at least as fast as $(h\bar h)^{(2-d)/2}$, which corresponds to the
exchange of the state of lowest dimension allowed by the unitarity bound.}
\[
R\simeq \partial \Gamma \ ,
\]%
to all orders in the coupling $g^{2}$.

In order to explore the consequences of the results of the previous
sections, we must analytically continue equation (\ref{generalex}) to find
the partial wave expansion of the Lorentzian amplitude $\hat{\mathcal{A}}=
\mathcal{A}^{\circlearrowleft }$. Using the perturbative form,
\[
\mathcal{A}(z,\bar{z})\simeq \sum \left( 1+\partial \Gamma \right) \left(
1+\Gamma \partial +\frac{1}{2}\Gamma ^{2}\partial ^{2}+\frac{1}{3!}\Gamma
^{3}\partial ^{3}+\cdots \right) \mathcal{S}_{h,\bar{h}}(z,\bar{z})\ ,
\]%
of equation (\ref{generalex}), we just need to compute the analytic
continuation
\[
\left[ \left( \partial _{h}+\partial _{\bar{h}}\right) ^{n}\,\mathcal{S}_{h,%
\bar{h}}(z,\bar{z})\right] ^{\circlearrowleft }\ .
\]%
This can be easily determined using the OPE expansion
\[
\mathcal{S}_{h,\bar{h}}(z,\bar{z})=z^{\frac{\Delta_1+\Delta_2}{2}-h}\,
\bar{z}^{\frac{\Delta_1+\Delta_2}{2} -\bar{h}%
}\sum_{n,\bar{n}\geq 0}z^{-n}\bar{z}^{-\bar{n}}c_{n,\bar{n}}(h,\bar{h}%
)\ \ \ +\ \ \ \left( z\leftrightarrow \bar{z}\right) \ ,
\]%
of the \emph{S}--channel partial waves around $z,\bar{z}\sim
\infty $ (see \cite{Paper2}). The differential operator
\[
\tilde{\partial} =  z^{-h}\bar{z}^{-\bar{h}}\,\partial \,z^{h}\bar{z}^{%
\bar{h}}=\partial +\ln (z\bar{z})\ ,
\]%
acting on $\mathcal{S}_{h,\bar{h}}$ for $h,\bar{h}\in (\Delta_1+\Delta_2)/2 +\mathbb{N}_{0}$%
, is invariant under the analytic continuation $\circlearrowleft $.
Therefore,
\begin{eqnarray*}
\left[ \partial ^{n}\mathcal{S}\right] ^{\circlearrowleft } &=&\left[ \left(
\tilde{\partial}-\ln (z\bar{z})\right) ^{n}\mathcal{S}\right]
^{\circlearrowleft } \\
&=&\left( \tilde{\partial}-\ln (e^{\,2\pi i}z\bar{z})\right) ^{n}\,\mathcal{S%
} \\
&=&\left( \partial -2\pi i\right) ^{n}\,\mathcal{S}\ .
\end{eqnarray*}%
The Lorentzian amplitude $\hat{\mathcal{A}}=\mathcal{A}^{\circlearrowleft }$
is then given by
\[
\hat{\mathcal{A}}(z,\bar{z})\simeq \sum \left( 1+\partial \Gamma \right)
\left( 1+\Gamma (\partial -2\pi i)+\frac{1}{2}\Gamma ^{2}(\partial -2\pi
i)^{2}+\frac{1}{3!}\Gamma ^{3}(\partial -2\pi i)^{3}+\cdots \right) \mathcal{S}_{h,\bar{h}}(z,\bar{z})\ .
\]
Focusing in the small $z,\bar{z}$ regime we can write
\[
\hat{\mathcal{A}}(z,\bar{z})\simeq \int dh d\bar{h}\left( 1-2\pi i\Gamma +
\frac{2\pi i}{2}(2\pi i+\partial )\Gamma ^{2}-\frac{2\pi i}{3!}(2\pi
i+\partial )^{2}\Gamma ^{3}+\cdots \right) \mathcal{I}_{h,\bar{h}}(z,\bar{z})\ ,
\]
where we have integrated by parts inside the integral over conformal
dimensions $h,\bar{h}$. In the large $h,\bar{h}$ limit we can neglect the
derivative $\partial  =  \partial _{h}+\partial _{\bar{h}}$ with respect
to the constant $2\pi i$, obtaining
\begin{equation}
\hat{\mathcal{A}}(z,\bar{z})\simeq \int dh d\bar{h}\,e^{-2\pi i\,\Gamma (h,\bar{h})}\,
\mathcal{I}_{h,\bar{h}}(z,\bar{z})\ .  \label{xyz22}
\end{equation}
Hence, in the impact parameter representation of the reduced Lorentzian amplitude $\hat{\mathcal{A}}$,
the anomalous dimensions $2\Gamma$ play the role of a phase shift.

\subsection{Impact Parameter Representation}

Now we wish to find an explicit form of the impact parameter representation for the
Lorentzian amplitude $\hat{A}$ in (\ref{zzz2}). First we recall a basic result derived in \cite{Paper2}. 
For $p ,\bar{p}$ in the past Milne wedge $-\mathrm{M}$, the impact parameter
partial wave $\mathcal{I}_{h,\bar{h}}$ admits the integral representation   \footnote{
The impact parameter representation derived in this section is valid in
general for $p=p_{3}$ and $\bar{p}=p_{2}$, with $p_{1}=p_{4}=0$. The general
case is then related by a conformal transformation, whose precise form is
rather cumbersome, but reduces to $p\simeq p_{3}-p_{1}$ and $\bar{p}\simeq
p_{2}-p_{4}$ for the case of interest $\left\vert p_{i}^a\right\vert \ll 1$.}
 over the future Milne wedge 
$\mathrm{M}$
\begin{eqnarray*}
\mathcal{I}_{h,\bar{h}} &=&\mathcal{N}_{\Delta _{1}}\mathcal{N}_{\Delta_{2}}
\left( -p ^{2}\right) ^{\Delta _{1}}\left( -\bar{p}^{2}\right)
^{\Delta _{2}}\int_{\mathrm{M}}
\frac{dx }{\left\vert x \right\vert^{d-2\Delta _{1}}}\,
\frac{d\bar{x}}{\left\vert \bar{x}\right\vert^{d-2\Delta _{2}}}\,
e^{-2p \cdot x -2\bar{p}\cdot \bar{x}} \\
&&\ \ \ \ \ \ \ \ \ \ \ \ \ \ \ \ \ \ \ \ \ \ \ 
4h\bar{h}~\delta \left( 2x \cdot \bar{x}+h^{2}+\bar{h}^{2}\right) 
\,\delta \left( x ^{2}\bar{x}^{2}-h^{2}\bar{h}^{2}\right) ~,
\end{eqnarray*}
where the cross ratios $z,\bar{z}$ are related to $p,\bar{p}$ as in (\ref{cba}) and 
the constant ${\cal N}_\Delta$ is given by
\[
\mathcal{N}_{\Delta }=\frac{2\pi ^{1-\frac{d}{2}}}{\Gamma \left( \Delta
\right) \Gamma \left( 1+\Delta -\frac{d}{2}\right)}=\frac{4\pi\, \mathcal{C}_{\Delta }}{\Gamma(\Delta)^2}\ .
\]
Expression (\ref{xyz22}) for the reduced amplitude becomes then
\begin{equation}
\hat{\mathcal{A}} \simeq \mathcal{N}_{\Delta _{1}}\mathcal{N}_{\Delta
_{2}}\left( -p ^{2}\right) ^{\Delta _{1}}\left( -\bar{p}^{2}\right)
^{\Delta _{2}}
\int_{\mathrm{M}}\frac{dx }{\left\vert x \right\vert^{d-2\Delta _{1}}}\,
\frac{d\bar{x}}{\left\vert \bar{x}\right\vert^{d-2\Delta _{2}}}
\,e^{-2p \cdot x -2\bar{p}\cdot \bar{x}}\,
e^{-2\pi i\,\Gamma( h ,\bar{h}) }\ ,
\label{impred}
\end{equation}
where $\Gamma(h,\bar{h})$ depends on $x,\bar{x}$ through
\begin{equation}
h^{2}\bar{h}^{2} = x ^{2}\,\bar{x}^{2} \ , \ \ \ \ \ \ \ \ \ \ \ \ \ \ \ \ \  
h^{2}+\bar{h}^{2} = -2x \cdot \bar{x} \ .
\label{abc1}
\end{equation}
The fact that $\hat{\mathcal{A}}$ is uniquely a function of
the cross--ratios $z,\bar{z}$, translates into the fact that the phase shift 
$\Gamma $ depends only on $x ^{2}\,\bar{x}^{2}$ and $-2x \cdot\bar{x}$.

To write the impact parameter representation for the full Lorentzian amplitude $\hat{A}$, 
consider first the boundary propagators in (\ref{zzz2}). For 
$p ,\bar{p}$ in the past Milne wedge $-{\rm M}$ we have
$$
\left( p ^{2}+i\epsilon _{p }\right) ^{-\Delta _{1}} = i^{2\Delta _{1}}\left( -p ^{2}\right) ^{-\Delta _{1}} \ ,
\ \ \ \ \ \ \ \ \ \ \ \ \ \ \ \ \ 
\left( \bar{p}^{2}-i\epsilon _{\bar{p}}\right) ^{-\Delta _{2}} =i^{-2\Delta _{2}}\left( -\bar{p}^{2}\right)^{-\Delta _{2}}\ ,
$$
where we recall that $\epsilon _{p }=\epsilon \func{sign}\left( -x_0\cdot p \right)$ with $x_0\in\mathrm{M}$.
Rotating the radial part of the $x,\bar{x}$ integrals over the Milne wedges in (\ref{impred}), so that
$x \to ix$ and $\bar{x}\to -i\bar{x}$,  (\ref{zzz2}) becomes
\begin{equation}
\hat{A}\left( p ,\bar{p}\right) \simeq (2\omega\, i)^{-2\Delta _{1}-2\Delta _{2}}
\mathcal{C}_{\Delta _{1}}\mathcal{C}_{\Delta _{2}}\mathcal{N}_{\Delta _{1}}
\mathcal{N}_{\Delta _{2}}\int_{\mathrm{M}}\frac{dx }{\left\vert x \right\vert
^{d-2\Delta _{1}}}\,\frac{d\bar{x}}{\left\vert \bar{x}\right\vert
^{d-2\Delta _{2}}}\,e^{2ip \cdot x -2i\bar{p}\cdot \bar{x}}\,e^{-2\pi i\,\Gamma ( h ,\bar{h})}\ .
\label{risultato}
\end{equation}
Although this representation was derived assuming $p ,\bar{p}$ in the past Milne wedge we claim it is valid for
generic $p ,\bar{p}\in \mathbb{M}^d$. In fact,
for the $\Gamma=0$  non--interacting amplitude,
we recover the boundary propagators from 
the Fourier transform (which we recall  in appendix 
\ref{app3})
\begin{equation}
\mathcal{N}_{\Delta }\int_{\mathrm{M}}\frac{dx }{\left\vert x
\right\vert ^{d-2\Delta }}~e^{\pm 2ip \cdot x }=\frac{1}{\left( p
^{2}\pm i\epsilon _{p }\right) ^{\Delta }}\ .
\label{suca1234}
\end{equation}

\subsection{Anomalous Dimensions of Double Trace Operators}\label{anomdim}

We are now in position to use the AdS/CFT prediction given by equations (\ref{eikEQ}) and (\ref{CFTamp})
to determine the phase shift  in the impact parameter representation (\ref{risultato})
and therefore to compute the anomalous dimension of double trace primary operators. First replace
(\ref{risultato}) in (\ref{CFTamp})
\begin{eqnarray*}
A_{eik}& \simeq &\omega ^{4}\,\left( 2\omega\, i\right) ^{-2\Delta _{1} -2\Delta _{2}}
\mathcal{C}_{\Delta _{1}}\mathcal{C}_{\Delta _{2}}
\mathcal{N}_{\Delta _{1}}\mathcal{N}_{\Delta _{2}}\\
&&\int dt_{1}\cdots
dt_{4}\,F\left( t_{1}\right) F\left( t_{2}\right) F^{\star }\left(
t_{3}\right) F^{\star }\left( t_{4}\right) \,e^{i\omega \left(
t_{3}-t_{1}\right) +i\omega \left( t_{4}-t_{2}\right) }\\
&&\int_{\mathrm{M}}\frac{dx }{\left\vert x \right\vert^{d-2\Delta _{1}}}\,
\frac{d\bar{x}}{\left\vert \bar{x}\right\vert^{d-2\Delta _{2}}}\,
e^{i\left( t_{3}-t_{1}\right) x_{0}\cdot x +i\left(t_{4}-t_{2}\right) x_{0}\cdot \bar{x}
- 2 i q\cdot x - 2i \bar{q}\cdot \bar{x}}\,e^{-2\pi i~\Gamma (h ,\bar{h} ) }\ .
\end{eqnarray*}
At high energy $\omega $, we have $t_{1}\sim t_{3}$ and $t_{2}\sim t_{4}$. Hence, the
integrals over the time sums $\frac{1}{2}\int d\left( t_{1}+t_{3}\right) F\left(
t_{1}\right) F^{\star }\left( t_{3}\right) $ and $\frac{1}{2}\int d\left(
t_{2}+t_{4}\right) F\left( t_{2}\right) F^{\star }\left( t_{4}\right) $ give
an overall factor of $2$ from the normalization (\ref{norm}). 
We are then left with the integrals over the differences, which give
\[
\left( 2\pi \right) ^{2}\,\delta \left( x_{0}\cdot x +\omega \right)
\delta \left( x_{0}\cdot \bar{x}+\omega \right) ~.
\]
It is easy to see that the integral in $x $ in the future Milne
wedge $\mathrm{M}$ at fixed time component $x_{0}\cdot x $ is
equivalent to the integral over points ${\rm w}$ in the hyperboloid $H_{d-1}$,
with the change of coordinates
\begin{eqnarray*}
x  &=&-\frac{ \omega }{x_{0}\cdot {\rm w}} \,{\rm w}   \ ,\\
\int_{\mathrm{M}}dx ~\delta \left( x_{0}\cdot x +\omega \right)
&=&2^{d}\,\omega ^{d-1}\int_{H_{d-1}}\frac{d{\rm w}}{\left( -2x_{0}\cdot {\rm w}\right)^{d}}\ .
\end{eqnarray*}
We then get
\begin{eqnarray}
A_{eik}& \simeq &2\left( 2\pi \omega \right) ^{2}\,i^{-2\Delta _{1}-2\Delta _{2}}\mathcal{C}
_{\Delta _{1}}\mathcal{C}_{\Delta _{2}}\mathcal{N}_{\Delta _{1}}\mathcal{N}
_{\Delta _{2}} \label{lastaeik}   \\
&&\int_{H_{d-1}}\frac{d{\rm w}}{\left( -2x_{0}\cdot {\rm w}\right) ^{2\Delta _{1}}}
\,\frac{d\bar{{\rm w}}}{\left( -2x_{0}\cdot \bar{{\rm w}}\right) ^{2\Delta _{2}}}
\,\exp\left( 2i \omega \,\frac{q\cdot {\rm w}}{x_{0}\cdot {\rm w}}+2i \omega \,\frac{\bar{q}\cdot \bar{{\rm w}}}{
x_{0}\cdot \bar{{\rm w}}}-2\pi i~\Gamma (h ,\bar{h}) \right) \ , \nonumber
\end{eqnarray}
where $\Gamma(h,\bar{h})$ depends on ${\rm w},\bar{{\rm w}}$ through
$$
4h\bar{h} = \frac{\left( 2\omega \right) ^{2}}{\left( x_{0}\cdot {\rm w}
\right) \left( x_{0}\cdot \bar{{\rm w}}\right) } \ ,
\ \ \ \ \ \ \ \ \ \ \ \ \ \ \ 
\frac{\bar{h}}{h}+\frac{h}{\bar{h}} = -2 {\rm w}\cdot {\rm \bar{w}} \ .
$$
We conclude that a double trace primary operator with large $h,\bar{h}$ can be described 
in AdS by two particles approximately following two null geodesics as in Figure \ref{geodesics}, with 
impact parameter $r=\log(h/\bar{h})$ and momenta ${\bf k}$ and  ${\bf \bar{k}}$ satisfying
$s= -2{\bf k} \cdot {\bf \bar{k}} = 4h\bar{h}$.

Finally, reverting equation (\ref{lastaeik}) to the embedding space notation, 
by replacing ${\rm w},\bar{{\rm w}},x_{0},2\omega q,$ $2\omega \bar{q}$ 
with $\mathbf{w,\bar{w},x}_{0}$, $\mathbf{q,\bar{q}}$,  we recover (\ref{eikEQ}), 
with a prediction for the large $h,\bar{h}$ behavior of the anomalous
dimensions due to the AdS exchange of a spin $j$ particle of
dimension $\Delta $,
\[
2\Gamma \left( h,\bar{h}\right) \simeq-\frac{g^{2}}{2\pi }\left( 4h\bar{h}\right)
^{j-1}~\Pi _{\perp }\left( h/\bar{h}\right) ~.
\]
The transverse propagator $\Pi _{\perp }$  is the Euclidean scalar propagator on $H_{d-1}$ with dimension $\Delta -1$.
Its explicit form in terms of the hypergeometric 
function is
\begin{eqnarray*}
\Pi _{\perp }(h,\bar{h}) &=& \frac{1}{2\pi^{\frac{d}{2}-1}}\,\frac{\Gamma
\left( \Delta -1\right) }{\Gamma\left( \Delta -\frac{d}{2} +1\right) }
\,\left( \frac{\left( h-\bar{h}\right) ^{2}}{h\bar{h}}\right) ^{1-\Delta }\  \\
&&F\left( \Delta -1,\frac{2\Delta -d+1}{2},2\Delta -d+1,~-\frac{4h\bar{h}}{%
\left( h-\bar{h}\right) ^{2}}\right) \ .
\end{eqnarray*}%
In particular, in dimensions $d=2$ and $d=4$ the above expression simplifies
to%
\begin{eqnarray*}
\Pi _{\perp }(h,\bar{h}) &=&\frac{1}{2\left( \Delta -1\right) }
\left( \frac{h}{\bar{h}}\right) ^{1-\Delta}\ \ \ \ \ \ \ \ \ \ \ \ \ \ \ \left(d=2\right)\ ,  \\
&=&\frac{1}{2\pi }\frac{{h}^{2}}{h^{2}-\bar{h}^{2}}
\left( \frac{h}{\bar{h}}\right) ^{1-\Delta}\ \ \ \ \ \ \ \ \ \ \ \ \ \ \left( d=4\right)\ .
\end{eqnarray*}

The anomalous dimensions just obtained are exactly the same  \footnote{In \cite{Paper1,Paper2} a different convention for the coupling constant $g_{%
\mathrm{here}}^{2}=4^{\,3-j}\,2\pi \,G_{\mathrm{there}}$ was used.} 
we obtained in \cite{Paper2}, where we only considered tree level
interactions based on a  shock wave computation in AdS \cite{Paper1}.
In other words, the loop corrections to the
anomalous dimensions of primary operators with large $h,\bar{h}$
are subleading with respect to the tree level contribution. This
is reminiscent of the flat space statement that the loop
corrections to the phase shift of large spin partial waves are
subleading with respect to the tree level contribution. We must therefore retract
the conjecture we put forward in \cite{Paper2}, which included
contributions from all orders in perturbation theory to the
anomalous dimensions in the large $h,\bar{h}$ limit. 

We emphasize that, for large $h,\bar h$, the anomalous dimensions are
dominated by the AdS particles with highest spin.
Moreover, when $h\gg\bar h$ the lightest particle of maximal spin
determines $\Gamma$, since in this limit the propagator $\Pi_\perp \sim (h/\bar h)^{1-\Delta}$.
In theories with a gravitational description, this particle is 
\emph{the graviton}. This yields a universal prediction for CFT's
with AdS duals in the gravity limit 
\be 
2\Gamma (h,\bar{h})\simeq -16Gh\bar{h}\,\Pi_{\perp }( h/\bar{h}) \ \ \ \ \ \ \ \ \ \ 
\left( h\sim\bar{h}\rightarrow \infty \ ,\ \ \ \ h\gg\bar h\, ,\ \ \ \ \ \ \Delta =d\right)\ ,
\label{sucabis}
\ee 
where $\Pi _{\perp }$ is the Euclidean
scalar propagator in $H_{d-1}$ with mass squared $d-1$.

Recall \cite{Paper2} that the impact parameter distance $r$ is given by $%
r=\ell \,\mathrm{ln}(h/\bar{h})$. Keeping in mind the canonical example of
the duality between strings on AdS$_{5}\times S^{5}$ and $\mathcal{N}=4$ SYM
theory, we expect (\ref{sucabis}) to be valid for large $r\gg \ell $.
Corrections to (\ref{sucabis}), due to massive KK modes of the graviton,
will start to be relevant at $r\sim \ell $. These corrections are computable
with an extension of the methods of this paper, which includes the sphere $S^{5}$ 
in the transverse space.
More complex, as in flat space, are the corrections due to string
effects \cite{ACV,GG}. As in flat space, particles of all
spins are exchanged, resulting in an effective reggeon interaction of spin
approximately $2$ for large string tension. 
As recalled in \cite{GG},
in flat space the leading corrections to the pure gravity result occur
due to tidal forces which
excite internal modes of the scattering strings. These effects start to be
relevant at impact parameters of the order of $\ell _{\mathrm{Plank}}\left(
\mathcal{E}\ell _{s}\right) ^{2/(d-1)}$, where $\ell _{\mathrm{Plank}}$ is
Planck length in the $\left( d+1\right) $--dimensional spacetime, and
where $%
\mathcal{E}$ is the energy of the process. Translating into AdS$_{5}$
variables, we then expect tidal string excitations to play a role at $%
r\lesssim G^{1/3}\ell ^{\,1/3}\ell _{s}^{\,2/3}(h\bar{h})^{1/3}$, i. e.  at 
$\mathrm{ln}(h/\bar{h})\lesssim (h\bar{h})^{1/3}N^{-2/3}\lambda ^{-1/6}$,
where $\lambda =(\ell /\ell _{s})^{4}$ is the 't Hooft coupling of the YM
theory. We shall discuss these effects extensively in a forthcoming
publication \cite{Paper4}.

\section{Future Work}    \label{futurework}

In this paper we have derived the eikonal approximation for high energy
interactions in Anti--de Sitter spacetime. We have been working uniquely in
the supergravity approximation, but we plan to extend these results by
including string effects \cite{Paper4}. Discussing, for concreteness, the
duality between strings on AdS$_{5}\times S_{5}$ and $\mathcal{N}=4$ SYM
theory, we shall address the following issues

\begin{itemize}
\item At large 't Hooft coupling $\lambda =\left( \ell /\ell
_{s}\right) ^{4} $, the leading correction to graviton exchange
will come from the contributions of the leading Regge trajectory.
The effective spin $j$ of the exchanged particle will now depend
on the transverse momentum transfer. This requires an extension of
Regge theory to conformal field theories which is quite natural in
our formalism, with results which reproduce and extend those of
\cite{PolStra}.

\item At weak coupling $\lambda $, high energy interactions are dominated by
Pomeron exchange. Following the initial results of \cite{PolStra}, we shall
relate our formalism to that of BFKL \cite{BFKL, LipatovSol, LipatovRev},
describing hard pomeron exchange at weak coupling, including the
non--trivial transverse dependence relevant at non--vanishing momentum
transfer.

\item The relation of phase shift and anomalous dimension suggests an
extension of the results of this paper to the weak coupling $\lambda
\rightarrow 0$ regime, following the ideas of Amati, Ciafaloni and Veneziano
\cite{ACV} on high energy string scattering. The phase shift $\Gamma $ will
become an operator acting on two--string states, which will include both an
orbital part as well as a contribution from the internal excitation of the
two scattering strings. A natural candidate for $\Gamma $ will be a
generalization, to double trace operators, of the dilatation operator \cite%
{Beisert} which has played a crucial role in analyzing the spectrum of
single trace states in $\mathcal{N}=4$ SYM theory.
\end{itemize}

\section*{Acknowledgments}
We would like to thank  N. Gromov, J. Maldacena, G. Veneziano and P. Vieira for discussions and comments.
Our research is supported in part by INFN, by the MIUR--COFIN contract 2003--023852,
by the EU contracts MRTN--CT--2004--503369, MRTN--CT--2004--512194, by the INTAS contract 03--51--6346,
by the NATO grant PST.CLG.978785 and by the FCT contract POCI/FP/63904/2005. 
LC is supported by the
MIUR contract \textquotedblleft Rientro dei cervelli\textquotedblright \ part VII.
JP is funded by the FCT fellowship SFRH/BD/9248/2002. \emph{Centro de F\'{\i}sica do Porto} is partially
funded by FCT through the POCTI program. MC and LC have been partially supported by the Galileo Institute
for Theoretical Physics, during the program \textit{String and M theory approaches to particle physics and
cosmology}, where part of this work was completed.

\vfill

\eject

\appendix


\section{General Spin $j$ Interaction}\label{app1}


We wish to extend to result $G({\bf w}\cdot{\bf \bar w})=\Pi_\perp({\bf k},{\bf \bar k})$, derived in section \ref{sectTP}, to the case of
general $j$. To this end we use equation (\ref{propeqBIS}), contracting both sides with
$$
(-2)^j\, {\bf k}_{\alpha_1}\cdots{\bf k}_{\alpha_j}\, {\bf \bar{k}}_{\beta_1}\cdots{\bf \bar{k}}_{\beta_j}
$$
and integrating against
$$
\int_{-\infty}^{\infty}du d\bar{v} = \frac{(2\omega)^2}{({\bf x}_0\cdot{\bf w})^2({\bf \bar{x}}_0\cdot{\bf \bar{w}})^2}\,\int_{-\infty}^{\infty}d\lambda d\bar{\lambda}\ .
$$
Using the explicit form of the $\delta$--function in the  $\{u,v, {\bf w} \}$ coordinate system give in section \ref{sectTP},
the RHS reduces to
\be
2i\,(2\omega)^{2j}\, \frac{(1+v\bar{u}/4)^{2j-2}}{
({\bf x}_0\cdot{\bf w})^{2j+2}}\,
\delta_{H_{d-1}}({\bf w},{\bf \bar{w}})\ .
\label{RHSbis}
\ee
Next we consider the LHS of (\ref{propeqBIS}). First we note that the non--vanishing components of the covariant derivatives of ${\bf k}$ are
given by
$$
\nabla_v {\bf k}_v=-\frac{\omega u}{2}\ ,\ \ \ \ \ \ \ \ \ \nabla_v {\bf k}_{\chi}=\nabla_{\chi} {\bf k}_v=\frac{\omega}{{\chi}}\ ,
$$
where we explicitly parametrize the metric on $H_{d-1}$ as in section \ref{sectTP}.
Using these facts, together with the explicit form of the metric and with
$$
\Box_{{\rm AdS}} {\bf k}_\alpha = -d\cdot {\bf k}_\alpha\ , \ \ \ \ \ \ \ \ \ \nabla_\gamma {\bf k}_\alpha\ ,
\nabla^\gamma {\bf k}_\beta = \frac{{\chi}^2-1}{{\chi}^2} {\bf k}_\alpha {\bf k}_\beta\, ,
$$
we conclude, after a tedious but straightforward computation, that
\beas
&&(-2)^j\, {\bf k}_{\alpha_1}\cdots{\bf k}_{\alpha_j}\, {\bf \bar{k}}_{\beta_1}\cdots{\bf \bar{k}}_{\beta_j}
\Box_{\rm AdS} \Pi_\Delta^{\alpha_1,\cdots,\alpha_j,\beta_1,\cdots,\beta_j}=\\
&&=\Box_{\rm AdS} \Pi_\Delta^{(j)}+j\,\left[2\frac{{\chi}^2-1}{{\chi}}\partial_{\chi}+(d+j-1)-\frac{j+1}{{\chi}^2}\right]\Pi_\Delta^{(j)} + \partial_u(\cdots)\\
&&=\left[\Box_{{\rm H}_{d-1}}+2(j+1)\frac{{\chi}^2-1}{{\chi}}\partial_{\chi}+j(d+j-1)-\frac{j(j+1)}{{\chi}^2}\right]\Pi_\Delta^{(j)} + \partial_u(\cdots)\, ,
\eeas
where we do not show the explicit terms of the form $\partial_u(\cdots)$ since they will vanish once integrated along the two geodesics.
Note that the terms in $\cdots$ contain also other components of the spin--$j$ propagator aside from $\Pi_\Delta^{(j)}$.
We conclude that (\ref{RHSbis}) must be equated to
$$
-2i\,(2\omega)^{2j}\,\left(1+\frac{v\bar{u}}{4}\right)^{2j-2} \left[\Box_{{\rm H}_{d-1}}-(\Delta+j)(\Delta-d-j)
+2(j+1)\frac{{\chi}^2-1}{{\chi}}\partial_{\chi}-\frac{j(j+1)}{{\chi}^2}
\right]
\frac{G({\bf w},{\bf \bar{w}})}{({\chi}\bar{{\chi}})^{j+1}}\ .
$$
Using the fact that
$$
[\Box_{{\rm H}_{d-1}},{\chi}^{-1-j}]=\frac{j+1}{{\chi}^{1+j}}\left(-2\frac{{\chi}^2-1}{{\chi}}\partial_{\chi}+(j-d+3)-\frac{j+2}{{\chi}^2}\right)\ ,
$$
we deduce again that
$$
\left[ \Box_{H_{d-1}} +1-d -\Delta( \Delta-d ) \right] G(  {\bf w} \cdot
{\bf \bar{w}} )
= -\delta(  {\bf w}, {\bf \bar{w}} )\ .
$$
and therefore the function $G$ is given by $\Pi_\perp$.


\section{Some Relevant Fourier Transforms}\label{app3}


Start by recalling  the standard generalized Feynman propagator%
\[
\frac{1}{\pi ^{d}}\int_{\mathbb{M}^d } \frac{dp }{\left( p ^{2}\mp i\epsilon \right)
^{\Delta }}~e^{2ix \cdot p }=\pm ~\frac{\pi ^{-\frac{d}{2}}\Gamma
\left( \frac{d}{2}-\Delta \right) }{\Gamma \left( \Delta \right) }~\frac{i}{%
\left( x ^{2}\pm i\epsilon \right) ^{\frac{d}{2}-\Delta }}~.
\]%
We now wish to consider the Fourier transform of interest%
\[
f\left( x \right) =\frac{1}{\pi ^{d}}\int_{\mathbb{M}^d } ~\frac{dp }{\left( p
^{2}-i\epsilon _{p }\right) ^{\Delta }}e^{2ix \cdot p }
\]%
We consider first the case $x ^{0} =  -x \cdot
x_{0}<0$. In this case $f\left( x \right) $ vanishes since we
can deform the $p ^{0}$ contour in the upper complex plane
$\func{Im}p ^{0}>0$. By Lorentz invariance, $f\left( x
\right) $ also vanishes whenever $x $ is spacelike, and
$f\left( x \right) $ is therefore supported only in the
future Milne wedge $\mathrm{M}$, where it is proportional to
$\left\vert x \right\vert ^{2\Delta -d}$. To find the
constant of proportionality, we note that, when $x ^{0}>0$ we
may deform the $p ^{0}$ contours in the lower complex plane and
show that
\[
f\left( x \right) =\frac{1}{\pi ^{d}}\int_{\mathbb{M}^d } \left[ \frac{dp }{\left(
p ^{2}+i\epsilon \right) ^{\Delta }}+\frac{dp }{\left( p
^{2}-i\epsilon \right) ^{\Delta }}\right] \, e^{2ix \cdot p }~.~\ \ \
\ \ \ \ \ \ \ \ \ \left( x ^{0}>0\right)
\]%
We then deduce that%
\begin{eqnarray*}
f\left( x \right)  &=&-i\,\frac{\pi ^{-\frac{d}{2}}\Gamma \left( \frac{d}{%
2}-\Delta \right) }{\Gamma \left( \Delta \right) }\left( i^{2\Delta
}-i^{-2\Delta }\right) \left\vert x \right\vert ^{2\Delta -d} \\
&=&\frac{2\pi ^{1-\frac{d}{2}}}{\Gamma \left( \Delta \right) \Gamma
\left( 1+\Delta -\frac{d}{2}\right) }\left\vert x \right\vert ^{2\Delta
-d}~\ \ \ \ \ \ \ \ \ \ \ \ \left( x \in \mathrm{M}\right)
\end{eqnarray*}%
and $f\left( x \right) =0$ for $x \notin \mathrm{M}$, thus proving
equation (\ref{suca1234}).


\vfill

\eject

\bibliographystyle{plain}

\begin{thebibliography}{99}

\bibitem{Paper1}
  L.~Cornalba, M.~S.~Costa, J.~Penedones and R.~Schiappa,
\textit{Eikonal approximation in AdS/CFT: From shock waves to four-pointfunctions},
\texttt{  [arXiv:hep-th/0611122]}.

\bibitem{Paper2}
L.~Cornalba, M.~S.~Costa, J.~Penedones and R.~Schiappa,
\textit{Eikonal Approximation in AdS/CFT: Conformal Partial--Waves and Finite N Four--Point Functions},
\texttt{[arXiv:hep-th/0611123]}.

\bibitem {Malda1}
J.~M.~Maldacena,
\textit{The Large $N$ Limit of Superconformal Field Theories and Supergravity}, Adv.\ Theor.\ Math.\ Phys.\ \textbf{2} (1998) 231,
\texttt{[arXiv:hep-th/9711200]}.

\bibitem {WittenGubser}
S.~S.~Gubser, I.~R.~Klebanov and A.~M.~Polyakov,
\textit{Gauge Theory Correlators from Non--Critical String Theory},
Phys.\ Lett.\ \textbf{B428} (1998) 105,
\texttt{[arXiv:hep-th/9802109]}.

\bibitem{w98}
E.~Witten,
\textit{Anti--de Sitter Space and Holography},
Adv.\ Theor.\ Math.\ Phys.\  {\bf 2} (1998) 253,
\texttt{[arXiv:hep-th/9802150]}.

\bibitem{Malda2}
O.~Aharony, S.~S.~Gubser, J.~M.~Maldacena, H.~Ooguri and Y.~Oz,
\textit{Large N Field Theories, String Theory and Gravity},
Phys.\ Rept.\ \textbf{323} (2000) 183,
\texttt{[arXiv:hep-th/9905111]}.

\bibitem{FreedmanRev}
E.~D'Hoker and D.~Z.~Freedman,
\textit{Supersymmetric Gauge Theories and the AdS/CFT Correspondence},
\texttt{[arXiv:hep-th/0201253]}.


\bibitem{LevySucher}
M.~Levy and J.~Sucher,
\textit{Eikonal Approximation in Quantum Field Theory},
Phys.\ Rev.\ \textbf{186} (1969) 1656.

\bibitem{tHooft}
G.~'t Hooft,
\textit{Graviton Dominance in Ultrahigh--Energy Scattering},
Phys.\ Lett.\ \textbf{B198} (1987) 61.



\bibitem{Paper4}
L.~Cornalba, M.~S.~Costa, J.~Penedones,
\textit{To appear}.


\bibitem{Jackiw}
  E.~Eichten and R.~Jackiw,
  \textit{Failure of the eikonal approximation for the vertex function in a boson
  field theory},
  Phys.\ Rev.\  D {\bf 4}, 439 (1971).


\bibitem{Kabat}
D.~Kabat and M.~Ortiz,
\textit{Eikonal Quantum Gravity and Planckian Scattering},
Nucl.\ Phys.\ \textbf{B388} (1992) 570,
\texttt{[arXiv:hep-th/9203082]}.\\
  D.~Kabat,
  \textit{Validity of the Eikonal approximation},
  Comments Nucl.\ Part.\ Phys.\  {\bf 20}, 325 (1992)
  \texttt{[arXiv:hep-th/9204103]}.





\bibitem{Osborn}
F.~A.~Dolan and H.~Osborn,
\textit{Conformal Partial Waves and the Operator Product Expansion},
Nucl.\ Phys.\ \textbf{B678} (2004) 491,
\texttt{[arXiv:hep-th/0309180]}.

\bibitem{Osborn22}
F.~A.~Dolan and H.~Osborn,
\textit{Conformal Four--Point Functions and the Operator Product Expansion},
Nucl.\ Phys.\ \textbf{B599} (2001) 459,
\texttt{[arXiv:hep-th/0011040]}.

\bibitem{fmmr98}
D.~Z.~Freedman, S.~D.~Mathur, A.~Matusis and L.~Rastelli,
\textit{Correlation Functions in the CFT$_d$/AdS$_{d+1}$ Correspondence},
Nucl.\ Phys.\ \textbf{B546} (1999) 96,
\texttt{[arXiv:hep-th/9804058]}.

\bibitem{kw99}
I.~R.~Klebanov and E.~Witten,
\textit{AdS/CFT Correspondence and Symmetry Breaking},
Nucl.\ Phys.\ \textbf{B556} (1999) 89,
\texttt{[arXiv:hep-th/9905104]}.





\bibitem{ACV}
D.~Amati, M.~Ciafaloni and G.~Veneziano,
\textit{Superstring Collisions at Planckian Energies},
Phys.\ Lett.\ \textbf{B197} (1987) 81.

\bibitem{GG}
  S.~B.~Giddings, D.~J.~Gross and A.~Maharana,
  \textit{Gravitational effects in ultrahigh-energy string scattering},
  \texttt{arXiv:0705.1816 [hep-th]}.

\bibitem{PolStra}
R.~C.~Brower, J.~Polchinski, M.~J.~Strassler and C.~I.~Tan,
\textit{The Pomeron and Gauge/String Duality},
\texttt{[arXiv:hep-th/0603115]}.\\
J.~Polchinski and M.~J.~Strassler,
\textit{Hard Scattering and Gauge/String Duality},
Phys.\ Rev.\ Lett.\ \textbf{88} (2002) 031601,
\texttt{[arXiv:hep-th/0109174]}.

\bibitem{BFKL} 
  M. T. Grisaru, H. J. Schnitzer and H. S. Tsao,
\textit{Reggeization of elementary particles in renormalizable gauge theories - vectors and spinors,}
  Phys. Rev.  D {\bf 8} (1973) 4498. \vspace{1mm}\newline
L. N. Lipatov, \textit{Reggeization of the Vector Meson and
the Vacuum Singularity in Nonabelian Gauge Theories, }Sov. J. Nucl. Phys.
\textbf{23} (1976) 642--656 \vspace{1mm}\newline
V. S. Fadin, E. A. Kuraev and L. N. Lipatov,\textit{\ On The Pomeranchuk
Singularity In Asymptotically Free Theories,} Phys. Lett. \textbf{B60}
(1975) 50--52 \vspace{1mm}\newline
E. A. Kuraev, L. N. Lipatov and V. S. Fadin, \textit{Multi--Reggeon
Processes in the Yang-Mills Theory, }Sov. Phys. JETP \textbf{44} (1976)
443--450\vspace{1mm}\newline
E. A. Kuraev, L. N. Lipatov and V. S. Fadin, \textit{The Pomeranchuk
Singularity in Nonabelian Gauge Theories, }Sov. Phys. JETP \textbf{45}
(1977) 199--204 \vspace{1mm}\newline
Ya. Ya. Balitsky and L. N. Lipatov,\textit{\ The Pomeranchuk Singularity in
Quantum Chromodynamics, }Sov. J. Nucl. Phys. \textbf{28} (1978) 822--829.

\bibitem{LipatovSol} L.~N.~Lipatov, \textit{The Bare Pomeron In Quantum
Chromodynamics}, Sov.\ Phys.\ JETP \textbf{63}, 904 (1986) [Zh.\ Eksp.\
Teor.\ Fiz.\ \textbf{90}, 1536 (1986)]. 

\bibitem{LipatovRev} L.~N.~Lipatov, \textit{Small--x physics in perturbative
QCD}, Phys.\ Rept.\ \textbf{286}, 131 (1997) \texttt{[arXiv:hep-ph/9610276]}%
. 

\bibitem{Beisert}
 N.~Beisert,
\textit{The dilatation operator of N = 4 super Yang-Mills theory and
 integrability},
Phys.\ Rept.\  {\bf 405}, 1 (2005)
\texttt{[arXiv:hep-th/0407277]}.




\end{thebibliography}

\end{document}